\documentclass[11pt,letterpaper]{article}
\usepackage[T1]{fontenc}
\usepackage[utf8]{inputenc}
\usepackage{lmodern}
\usepackage[letterpaper,margin=1in]{geometry}
\usepackage{amsmath,amssymb,amsthm,mathtools,thmtools}
\usepackage{microtype,booktabs,array,needspace}
\usepackage{color,xcolor,enumitem}
\usepackage[colorlinks=true,linkcolor=purple,citecolor=blue]{hyperref}
\usepackage[nameinlink]{cleveref}
\allowdisplaybreaks[2]
\newtheorem{introthm}{Theorem}
\crefname{introthm}{theorem}{theorems}
\Crefname{introthm}{Theorem}{Theorems}
\newtheorem{theorem}{Theorem}[section]
\newtheorem{lemma}[theorem]{Lemma}
\newtheorem{proposition}[theorem]{Proposition}
\newtheorem{corollary}[theorem]{Corollary}
\theoremstyle{definition}
\newtheorem{definition}[theorem]{Definition}

\theoremstyle{remark}

\BeforeBeginEnvironment{introthm}{\ifnum\value{introthm}=0
  \Needspace{14\baselineskip}\else\Needspace{5\baselineskip}\fi}
\BeforeBeginEnvironment{lemma}{\Needspace{5\baselineskip}}
\newcommand{\SM}{\operatorname{SM}}
\newcommand{\PM}{\operatorname{PM}}

\newcommand{\past}{\mathord\downarrow}
\newcommand{\future}{\mathord\uparrow}
\newcommand{\ind}{\mathbf 1}
\title{A $2.81^n$ upper bound	on the number of stable matchings}
\author{Ebrahim Ghorbani\thanks{\raggedright Hamburg University of Technology, Institute for Algorithms and Complexity, Hamburg, Germany. \texttt{ebrahim.ghorbani@tuhh.de}}\and  Matthias Mnich\thanks{\raggedright Hamburg University of Technology, Institute for Algorithms and Complexity, Hamburg, Germany. \texttt{matthias.mnich@tuhh.de}}}
\date{}
\begin{document}
\hypersetup{pageanchor=false}
\begin{titlepage}
\maketitle
\begin{abstract}
  We study the maximum number $\SM(n)$ of stable matchings in an instance with $n$ men and $n$ women. Karlin, Oveis Gharan, and Weber (STOC~2018) proved the first single-exponential upper bound; their argument gives $\SM(n)\le 2^{17n}$ for sufficiently large $n$.
  Palmer and P\'alv\"olgyi (FOCS~2021) improved this bound to $3.55^n+O(1)$.
  
 We prove that $\SM(n)<2.81^n$.
This brings the exponential base closer to $2.28$, the base
of the best known lower bound.
  More precisely, we bound the number of stable matchings in each instance in terms of each
person's number of stable partners, where a stable partner is someone matched to that person in at least one stable matching.
  Our proof assigns each person a weight determined by this number and uses these weights to bound the number of stable matchings by induction.
\end{abstract}
\thispagestyle{empty}
\end{titlepage}
\hypersetup{pageanchor=true}
\setcounter{page}{1}

\section{Introduction}
\label{sec:introduction}
A \emph{stable matching instance}, or simply an \emph{instance}, consists of $n\ge1$ men and $n$ women, each having a strict ranking of all members of the other set.
We call these rankings \emph{preferences}.
Equivalently, an instance is a complete bipartite graph with $n$ vertices in each class, where each vertex ranks the vertices in the opposite class.
A matching is a bijection between the two sets.
A man and a woman who are not matched with each other form a \emph{blocking pair} if each prefers the other to their assigned partner.
A matching is \emph{stable} if it has no blocking pair.

Gale and Shapley~\cite{GS} proved that every instance has a stable matching and gave an efficient algorithm to find one; see also the standard textbook by Gusfield and Irving~\cite[Chapter~1]{GI}.
Since their introduction in 1962, stable matching models have found applications in college admissions~\cite{GS} and the assignment of medical graduates to hospitals~\cite{Roth1984}.
Related matching models have also been used to organize kidney exchange~\cite{RSU2004}.
Galvin~\cite{Galvin1995} used the existence of stable matchings to resolve the Dinitz conjecture, which for many years was considered one of the widest-known open questions on list coloring.

Let $s(I)$ be the number of stable matchings of an instance $I$, and let $\SM(n)$ be the maximum of $s(I)$ over all instances with $n$ men and $n$ women.
Determining $\SM(n)$ is a long-standing problem, posed by Knuth in 1976~\cite{Knuth}.
Besides its interest as a counting problem, bounds on $\SM(n)$ give bounds on the time needed to enumerate all stable matchings~\cite{Gusfield1987}.
Enumeration also solves optimization problems such as sex-equal stable marriage~\cite{Kato1993} and balanced stable marriage~\cite{GRSZ2021}, by evaluating each stable matching and choosing the best one.
Bounds on $\SM(n)$ also give run time bounds for algorithms for the popular roommates problem~\cite{Kavitha2025}.

\paragraph{Lower bounds.} A simple construction gives $\SM(n)\ge2^{\lfloor n/2\rfloor}$.
Irving and Leather~\cite{IL} improved this to $\Omega(2.28^n)$ when $n$ is a power of two.
Thurber~\cite{Thurber} extended their construction to all $n$, obtaining a lower bound of the form $2.28^n/c^{\ln n}$ for an absolute constant $c>1$.

\paragraph{Upper bounds.} The immediate upper bound is $\SM(n)\le n!$, since there
are only $n!$ matchings.
The problem of improving this bound was already raised by Knuth~\cite{Knuth} in 1976.
In their by-now classic book on stable matchings, Gusfield and Irving~\cite[Section 1.3.2]{GI} list 12 open problems around stable marriage; one of these problems is to determine $\SM(n)$.
Stathopoulos~\cite{Stathopoulos2011} proved that $\SM(n)=O(n!/c^n)$
for some constant $c>1$. Drgas-Burchardt and \'Switalski~\cite{DS2013}
gave an upper bound of the form $(3/4+O(1/n))n!$.
The question of discovering $\SM(n)$ was re-iterated in the fundamental book on algorithmics of matching under preferences by Manlove~\cite[Section 2.2]{Manlove2013}.
Despite considerable work on the problem, the first upper bound of the form $c^n$, for an absolute constant $c$, was only obtained in 2018, by Karlin, Oveis Gharan, and Weber~\cite{KGW}.
Their argument gives $\SM(n)\le2^{17n}$ for sufficiently large $n$. 
The authors noted that their method does not give an exponential base close to the known lower bound of $2.28$.
Thus, they highlighted determining the precise asymptotic growth of $\SM(n)$ as a challenging open problem~\cite[Section~5]{KGW}.
Further, Cechl{\'a}rov{\'a}, Cseh, and Manlove~\mbox{\cite[Section~3.2.1]{CCM2019}} explicitly list improving the known bounds on $\SM(n)$ as an open problem in their survey.
Palmer and P{\'a}lv{\'o}lgyi~\cite{PP} thereafter proved that $\SM(n)\le3.55^n+O(1).$

\paragraph{Main contribution.} Our main result gives the best upper bound on the number of stable matchings.
By showing $\SM(n) \leq 2.81^n$, we bring the exponential base close to the known lower bound of $2.28$.
More precisely, we bound $s(I)$ in terms of the numbers of stable partners of the individual men and women.
A \emph{stable partner} is a partner occurring in at least one stable matching.
As any instance has a stable matching, each man and woman has between one and $n$ stable partners.

\begin{introthm}
\label{thm:main}
  There is a sequence $h_1,h_2,\ldots$ of positive numbers with $\sum_{d=1}^{\infty}h_d<\ln2.81$ such that the following holds.
  Let $I$ be an instance with $n$ men and $n$ women, and let $D_v^m$ and $D_v^w$ be the numbers of stable partners of man $v$ and woman $v$, respectively.
  Then
  \begin{equation*}
    s(I)\le \exp\left(\frac12\sum_{v=1}^{n} \left(\sum_{d=1}^{D_v^m-1}h_d +\sum_{d=1}^{D_v^w-1}h_d\right)\right) \enspace .
  \end{equation*}
  In particular,
  \begin{equation*}
    \SM(n)\le \exp\left(n\sum_{d=1}^{n-1}h_d\right)<2.81^n \enspace .
  \end{equation*}
\end{introthm}

\paragraph{Consequences of \Cref{thm:main}.}
Beyond an object worthwhile being studied by itself, our improved upper bound on $\SM(n)$ has combinatorial and algorithmic consequences.

Combinatorially, first, the bound of \Cref{thm:main} transfers to an upper bound
on the number of stable assignments in the stable roommates problem, in which the graph of an instance~$I$ need not be bipartite as in the stable matching problem.
Each person then ranks the other $n - 1$ persons.
As before, persons are matched in pairs, and such matching is stable if no two unmatched persons prefer each other over their current match.
As observed by Karlin, Oveis Gharan and Weber~\cite{KGW}, a standard reduction~\cite{DeanM2010} can map $I$ to a stable matching instance $I'$ with $n$ men and $n$ women, such that the stable roommate assignments are a subset of the stable matchings of $I'$.
We thus get:
\begin{corollary}
  Any stable roommates instance with $n$ persons and strict
  preferences has at most~$2.81^n$ stable matchings.
\end{corollary}
Second, from $s(I)$ we can derive an upper bound on the number $\textnormal{pop}(I)$ of so-called ``popular matchings'' of an instance $I$.
The notion of popular matchings dates back to G{\"a}rdenfors~\cite{Gardenfors1975} in~1975.
A matching $\mu$ is \emph{popular} if, for every other matching $\mu'$, one has $\bigl|\{v: v\text{ prefers }\mu\text{ to }\mu'\}\bigr|\ge\bigl|\{v: v\text{ prefers }\mu'\text{ to }\mu\}\bigr|$.
Any stable matching is popular, so $\textnormal{pop}(I)\geq s(I)$.
Conversely, popular matchings of~$I$ with men set $M$ and woman set $W$ can be mapped to stable matchings of an instance~$I'$, via a construction by Cseh and Kavitha~\cite{CsehK2018}.
Instance $I'$ has $2n$ men $M' = \{m_0,m_1: m\in M\}$ and $2n$ women $W'=W\cup\bigl\{d(m): m\in M\bigr\}$.
Each woman's ranks all acceptable $m_1$ men first, in her original order, followed by the $m_0$ men in the same order.
For each~$m_0$ its preference list places first the original women in $m$'s list, followed by $d(m)$; for $m_1$ start with $d(m)$ and then the women in $m$'s list; for $d(m)$ put $m_0$ before~$m_1$.
There is a projection $P$ from the stable matchings of $I'$ to the dominant matchings of $I$, obtained by deleting the dummy edges and replacing $(m_i,w)$ by $(m,w)$.
Here, a popular matching is \emph{dominant} if it is more popular than any (not necessarily popular) matching of larger cardinality; let $\textnormal{dom}(I)$ be the number of dominant matchings.
Cseh and Kavitha~\cite{CsehK2018} show $P$ to be surjective, so every dominant matching of $I$ is the projection of some stable matching of $I'$. 
So $\textnormal{pop}(I)=\textnormal{dom}(I)\leq s(I')$.
Every stable matching~$\mu'$ of $I'$ projects to a popular matching~$P(\mu')$ of the complete instance; the projection is therefore perfect.
And every dummy woman~$d(m)$ is matched in any stable matching of $I'$, otherwise $(m_1,d(m))$ would block, since $d(m)$ is $m_1$'s top choice. 
Thus every stable matching of~$I'$ is perfect.
We can complete the preference lists of $I'$ by putting all previously unacceptable partners below every original acceptable partner, in arbitrary strict order.
For the resulting instance~$\tilde{I'}$, every stable matching of $I'$ remains stable in $\tilde I'$: it is perfect and each new edge is worse at both endpoints than their current acceptable partners. 
Hence, $s(I') \leq s(\tilde I') \leq \SM(2n)$.
So $\textnormal{pop}(I) \leq \SM(2n)$.
Defining $\PM(n)$ as the maximum of $\textnormal{pop}(I)$ over all instances with $n$ men and $n$ women shows $\PM(n)\leq \SM(2n)$, so \Cref{thm:main} gives:
\begin{corollary}
  Any popular matching instance with $n$ men and $n$ women has at most \mbox{$\PM(n) \leq 7.90^n$} popular matchings.
\end{corollary}

Algorithmically, our bound on $\SM(n)$ reduces the run time $O\bigl(\SM(n) + 2\sqrt{\SM(n)}\bigl)$ of Kavitha's algorithm~\cite{Kavitha2025} for solving the popular roommates problem with strict preferences, from $O(7.32^n)$ to $O(6.17^n)$.
Further, for many practical applications of stable matchings the assumption of having complete preference lists for each man and each woman, is too strong, say for instance because of lack of information about all persons of the opposite sex.
Other applications demand relaxations on the structure of preference lists, which lead to relaxations of the notion of stability.
Among those, Miyazaki and Tamaki~\cite{MiyazakiT2026} consider the {\sc Max-SMTI} problem in which preference lists can be incomplete, and need not be strict but can contain ties.
A \emph{partial} matching $\mu$ is then a bijection between a subset of men and women.
In their work, such a partial  matching $\mu$ is called \emph{weakly stable} if there is no edge $(m,w)$ in the bipartite men-women graph, such that either both $m,w$ are unmatched in $\mu$ (and thus $(m,w)$ can be added to $\mu$ to give a larger matching), or both $m,w$ are matched in $\mu$ but each prefers the other to their assigned
partner.
The objective in {\sc Max-SMTI} is to compute a weakly stable matching of maximum cardinality.
They give an algorithm~\cite{MiyazakiT2026} to solve {\sc Max-SMTI} in time $O^*\left(\Bigl(1+\sqrt{4\cdot \SM(n)}\Bigr)^{2n}\right)$.
Using our bound on $\SM(n)$ from \Cref{thm:main} within their algorithm, we thus improve its run time from $O(22.74^n)$ as follows:
\begin{corollary}
  For any instance with $n$ men and $n$ women, whose preference lists may be incomplete or contain ties, a maximum-cardinality weakly stable matching can be found in time~$O(18.95^n)$.
\end{corollary}

\subsection{Related works}
Our result bounds the number of stable matchings in every instance.
Other work studies how this number can be computed, how it behaves under random preferences, and how many stable partners each man or woman can have.

\paragraph{Computing the number of stable matchings.}
Irving and Leather~\cite{IL} proved that counting stable matchings exactly is $\#\mathrm{P}$-complete.
For approximate counting, Dyer, Goldberg, Greenhill, and Jerrum~\cite{DGGJ2004} established an equivalence with $\#\mathrm{BIS}$, the problem of counting independent sets in a bipartite graph.
The reductions in both directions preserve the approximation.
For the setting when all men have the same ``master'' preference list over all women (or vice-versa), Irving, Manlove and Scott~\cite{IrvingMS2008} showed how to find the unique stable matching greedily in linear time.

However, restrictions on preferences do not always make counting easier.
Chebolu, Goldberg, and Martin~\cite{CGM2012} proved the same equivalence with $\#\mathrm{BIS}$ when preferences are determined by linear scores on three attributes, and when they are determined by Euclidean distances in two dimensions.
These results use the representation of stable matchings as downsets of the rotation poset~\cite{IL}, which we also use in our proof.

Sampling stable matchings presents related difficulties.
Bhatnagar, Greenberg, and Randall~\cite{BGR2008} examined a natural Markov chain that moves along the edges of the stable matching lattice.
They constructed instances for which its mixing time is exponential, even when rankings are determined by only two attributes.
They obtained the same conclusion for the range-five model, where positions of any given man or woman across preference lists lie within five consecutive ranks.
Their constructions also show that these restrictions allow exponentially many stable matchings.

\paragraph{The number of stable matchings in random instances.}
The choice of a probability model has a substantial effect on the number of stable matchings.
For independent, uniformly random complete preference lists, Pittel~\cite{Pittel1989} proved that the expected number of stable matchings is asymptotic to $e^{-1}n\ln n$.
By Markov's inequality, the number of stable matchings is at most $n(\ln n)^2$ with probability tending to one.
In particular, it is polynomial in $n$ with probability tending to one.

A common preference order can produce a different result.
In the Mallows model~\cite{Mallows1957}, the probability of a ranking $\pi$ is proportional to
$q^{\operatorname{inv}(\pi)}$, where $0<q<1$ and $\operatorname{inv}(\pi)$ counts inversions relative to a fixed reference order.
Hoffman, Levy, and Mossel~\cite{HLM2023} proved that, when preference lists are sampled independently from this model with a common reference order, the number of stable matchings is exponential in $n$ with probability tending to one, for each fixed $q$.
This work first appeared in Levy's thesis~\cite{Levy2017}.



\paragraph{The number of stable partners.}
Our bound also depends on the number of stable partners of each man and woman.
For independent, uniformly random complete preferences, Knuth, Motwani, and Pittel~\cite{KMP1990} showed that any fixed woman has $\Theta(\ln n)$ stable partners
with probability tending to one.
The same statement holds for any fixed man by symmetry.

Short preference lists can greatly reduce the number of stable partners.
Immorlica and Mahdian~\cite{IM2005} allowed women to have arbitrary complete preference lists, while each man independently chooses a uniformly random list of $k$ acceptable women.
For fixed $k$, the expected number of women with more than one stable partner is bounded by a constant depending only on $k$.

Unequal numbers of men and women can have a similar effect even with complete lists.
Ashlagi, Kanoria, and Leshno~\cite{AKL2017} considered $n$ men and $n+1$ women with independent, uniformly random complete preferences.
They proved that, with probability tending to one, all but a vanishing fraction of the men and women have a unique stable partner.
\medskip

For broader treatments of stable matching, we refer to Roth and Sotomayor~\cite{RS1990}, Gusfield and Irving~\cite{GI}, Manlove~\cite{Manlove2013}, and Knuth~\cite{Knuth1997}.
Roth~\cite{Roth2015} discusses the design of matching markets and their practical uses.

\subsection{Overview of the proof}
Let $I$ be a stable matching instance.
A rotation is a cyclic exchange of partners that changes one stable matching into another. Some rotations must happen before others; these requirements define a partial order on the rotations, giving the \emph{full rotation poset} $Q_{\rm f}(I)$.
Stable matchings correspond to the downsets of~$Q_{\rm f}(I)$, cf. Irving and Leather~\cite{IL}, which are sets that are closed from below in~$Q_{\rm f}$.
Write $J(P)$ for the number of downsets of a poset $P$.

\paragraph{Normalized instances.}
For the instance $I$, we move each person's stable partners to the front of their preference list, preserving their relative order, and call the resulting instance $\widehat I$ the \emph{normalization} of $I$.
We do this because our proof benefits from having the rotation order generated by shared participants, a property not assumed in the earlier exponential bounds of Karlin, Oveis Gharan, and Weber~\cite{KGW} and Palmer and P\'alv\"olgyi~\cite{PP}.
This means that the order is determined by each person's sequence of rotations.
For an arbitrary instance, these two orders need not coincide, but they coincide for normalized instances.
Furthermore, every stable matching of $I$ remains stable in $\widehat I$, and everyone's stable partners stay the same.
Thus, $s(I)\le s(\widehat I)=J(Q_{\rm f}(\widehat I))$.
Working with normalized instances therefore allows us to assume that the entire order of $Q_{\rm f}$ is generated by shared participants.

\paragraph{The branching potential.}
Let $\mathcal F$ be a family of finite posets containing $\varnothing$.
A nonnegative function $F$ on $\mathcal F$ with $F(\varnothing)=0$ is a \emph{branching potential} if, for every nonempty $Q\in\mathcal F$, there exist $Q_1,Q_2\in\mathcal F$ with $|Q_i|<|Q|$ such that $J(Q)=J(Q_1)+J(Q_2)$ and
\begin{equation}\label{eq:def-branching-potential}
  e^{F(Q_1)}+e^{F(Q_2)}\le e^{F(Q)} \enspace .
\end{equation}
Such a function gives the bound $J(Q)\le e^{F(Q)}$ by induction on $|Q|$:
The base case follows from $J(\varnothing)=1=e^{F(\varnothing)}$, and the induction step gives
\begin{equation}
\label{eq:branching-potential}
  J(Q)   = J(Q_1)+J(Q_2)  \le e^{F(Q_1)}+e^{F(Q_2)}
       \le e^{F(Q)} \enspace .
\end{equation}
This is a direct induction on the branching recurrence; compare the measure-and-conquer method of Fomin, Grandoni, and Kratsch~\cite{FGK}.


Let $\mathcal Q$ be the family of rotation posets of normalized instances
on the same $n$ men and $n$ women. We give each woman her partner's label in the man-optimal matching, whose existence and uniqueness follow from Gale and Shapley~\cite{GS}. For a given $Q\in\mathcal Q$, a person's \emph{degree}
is the number of rotations in $Q$ containing them,
which equals their number of stable partners minus one.
Write $d_v^m,d_v^w$ for the degrees of the man and woman
with label $v$, respectively.
Our goal is to find a branching potential on $\mathcal Q$
that depends on these degrees.
For a given $Q\in\mathcal Q$, the smaller posets $Q_1,Q_2$ are obtained by deleting certain rotations from $Q$.
These smaller posets involve the same people, but each person may belong to fewer rotations.
A bound depending only on $n$ would therefore not decrease.
Instead, we give a person of degree $d$ a contribution $W(d)$, with $W(0)=0$.
We choose $W$ to be increasing and bounded,
with decreasing increments $W(d)-W(d-1)$.
When a person's degree drops from $d$ to $d-1$, their weight
drops by this increment. Thus, deleting one rotation gives
a larger decrease when the person's degree is smaller.
This allows small extra weights at large degrees while
keeping larger decreases at small degrees.

We start with an unknown sequence $h_1\ge h_2\ge\cdots>0$, whose terms represent the increments $W(r)-W(r-1)$, and define $W(d)=\sum_{r=1}^d h_r$, and
\begin{equation*}
  F_h(Q)=\frac{F_m(Q)+F_w(Q)}2,
\end{equation*}
where $F_s(Q)=\sum_{v=1}^n W(d_v^s)$ for $s=m,w$.
We choose the increments $h_r$ so that $F_h$ is a branching potential.
Sufficient conditions for this are collected in \Cref{def:admissible}, where a sequence satisfying these conditions is called \emph{admissible}.
\begin{introthm}
\label{thm:admissible-potential}
  For every admissible sequence $(h_r)_{r\ge1}$, the function $F_h$ is a branching potential on $\mathcal Q$.
\end{introthm}

If an admissible sequence has finite sum $S=\sum_{r=1}^{\infty}h_r$, then $F_h(Q)\le nS$.
We construct such a sequence by solving a finite convex optimization problem that minimizes $S$ subject to conditions ensuring admissibility.
We obtain the following result.

\begin{introthm}
\label{thm:sum-admissible}
  There exists an admissible sequence $(h_r)_{r\ge1}$ such that $\sum_{r=1}^{\infty}h_r<\ln 2.81$.
\end{introthm}

Together, \eqref{eq:branching-potential},
\Cref{thm:admissible-potential,thm:sum-admissible} essentially prove
\Cref{thm:main}, since
\begin{equation*}
  s(I)\le e^{F_h(Q)}\le e^{nS}<2.81^n \enspace .
\end{equation*}

Our approach differs from those of Karlin, Oveis Gharan, and Weber~\cite{KGW} and Palmer and P\'alv\"olgyi~\cite{PP} in two ways.
First, we use a somewhat different rotation poset, whose order is generated entirely by shared participants.
Second, Karlin, Oveis Gharan, and Weber count downsets by choosing elements with many predecessors and successors, while Palmer and P\'alv\"olgyi reveal downsets along the participant chains in random order.
Our counting argument instead uses a potential depending on each person's degree to measure progress when rotations are deleted.

\medskip

The main technical ingredient is \Cref{thm:admissible-potential}.
In the rest of this subsection, we outline its proof.

\paragraph{Branch losses.}
Let $Q\in\mathcal Q$ and $\rho\in Q$.
Let $\past\rho$ consist of $\rho$ and all its predecessors in $Q$, and let~$\future\rho$ consist of~$\rho$ and all its successors.
Set $Q_1=Q\setminus\past\rho$ and $Q_2=Q\setminus\future\rho$.
The downsets of $Q$ can be partitioned into those containing $\rho$ and those avoiding $\rho$. This gives $J(Q)=J(Q_1)+J(Q_2)$.
We prove that $Q_1,Q_2\in\mathcal Q$ and that, in $Q_1,Q_2$, the degree of a person $v$ equals their degree in $Q$ minus the number of deleted rotations involving $v$.
Set $p^*(\rho)=F_h(Q)-F_h(Q_1)$ and $q^*(\rho)=F_h(Q)-F_h(Q_2)$, and put $\lambda(t)=-\ln(1-e^{-t})$ for $t>0$.
To show that $F_h$ is a branching potential, we need to
verify~\eqref{eq:def-branching-potential}.
For this, we seek a rotation $\rho$ for which
\begin{equation*}
  e^{-p^*(\rho)}+e^{-q^*(\rho)}\le1,
  \quad\text{equivalently}\quad
  \lambda(p^*(\rho))\le q^*(\rho) \enspace .
\end{equation*}
Let $V=\{v:d_v^m>0\}=\{v:d_v^w>0\}$ be the active labels.
The man and woman with label $v$ remain matched until their common first rotation $b_v$.
We prove that
\begin{equation}
\label{eq:summed-criterion-intro}
\tag{BC}
  \sum_{v\in V}\lambda(p^*(b_v))\le\sum_{v\in V}q^*(b_v).
\end{equation}
Then some $v\in V$ satisfies
$\lambda(p^*(b_v))\le q^*(b_v)$, so $\rho=b_v$
gives the required rotation.
We call $\lambda(p^*(b_v))$ the \emph{cost} of label $v$.

The proof plan is now more concrete: we first rewrite the
right-hand side of \eqref{eq:summed-criterion-intro} as a sum
of contributions from each person's rotations, and then show
that these contributions cover the total cost of all labels.

\paragraph{Grouping rotations.}
We first estimate the right-hand side of \eqref{eq:summed-criterion-intro}.
The \emph{support} of a rotation~$\rho$~is
\begin{equation*}
  S(\rho)=\{v\in V:b_v\preceq\rho\} \enspace .
\end{equation*}
It records which initial pairs must have changed partners by the time $\rho$ is performed. After a person's $r^{\textnormal{th}}$ rotation, they have had $r+1$ distinct partners, so the support of that rotation contains at least $r+1$ labels.
Consider the potential for men alone, and write $q(\rho)=F_m(Q)-F_m(Q\setminus\future\rho)$
for the decrease in the men's potential when $\rho$ and all its successors are deleted.
If $b_v\preceq\rho$, deleting~$\future b_v$ also deletes $\rho$.
Fix a man $u$, and let $\rho$ be his $r^{\textnormal{th}}$ rotation.
For each label $v\in S(\rho)$, deleting $\future b_v$ removes $\rho$ and all later rotations involving $u$.
The decrease in $u$'s contribution therefore includes $h_r$.
There are at least $r+1$ such labels $v$, so this same increment $h_r$ appears in at least $r+1$ terms of the sum $\sum_{v\in V}q(b_v)$.
This gives
\begin{equation}\label{eq:sum_q_v}
  \sum_{v\in V}q(b_v)=\sum_{u\in V}H_{d_u}+\Omega,
\end{equation}
where $H_d=\sum_{r=1}^d(r+1)h_r$ and $\Omega\ge0$ collects the extra contributions from supports larger than $r+1$.
The same identity holds for women.
We use these amounts to cover the costs on the left-hand side of~\eqref{eq:summed-criterion-intro}.
We need $\sum_r h_r$ to be finite so that $W(d)$ stays
bounded. We also need $H_d$ to grow fast enough to cover
terms such as $\lambda(2h_d)$.
For weights behaving like $C/r^\alpha$, the first requirement
needs $\alpha>1$, while the second rules out $\alpha>2$.
We choose $\alpha=2$, the largest possible exponent in
this family.
With this choice, the factor $r+1$ in the definition of
$H_d$ makes each term behave like $C/r$.
Thus $H_d\sim C\ln d$, while $\lambda(2h_d)\sim2\ln d$.
Choosing $C>2$ therefore makes $H_d$ eventually exceed
$\lambda(2h_d)$, giving the required comparison for large
degrees.

\paragraph{Considering men and women together.}
To prove \eqref{eq:summed-criterion-intro}, we use \eqref{eq:sum_q_v} to cover the total cost of all labels.
We estimate the costs for men and women separately.
For each person, we reserve~$H_{d_v}$ together with part of their first two rotations'  contributions to $\Omega$, and denote this amount by~$\beta_v$.
Roughly, these amounts cover the costs after we group labels according to their common first rotation.
The main difficulty comes from the following situation.

Start from the man-optimal matching.
Men $u$ and $v$ first exchange partners.
Then men $u$ and $x$ exchange partners. This rotation is a \emph{swap} (a rotation in which two men exchange partners).
It is $u$'s second rotation and $x$'s first, and its support is $\{u,v,x\}$.
In this situation, we call $x$ \emph{exceptional}.
We estimate the costs of these three labels together.
After the second swap, man $x$ is matched to woman $v$.
He has changed partners once, while she has changed partners twice.
They either stay together or both participate in another rotation.
Hence,
\begin{equation*}
  d_x^m=1\quad\Longleftrightarrow\quad d_v^w=2 \enspace .
\end{equation*}
When we express the cost bounds in terms of the reserved amounts $\beta_v$, the first case adds an extra~$h_1$ on the men's side, while the second saves $h_1$ on the women's side.
Since the cases occur together, these terms cancel when we combine the estimates.

Some costs may still remain uncovered when exceptional people change partners again. We examine their second and third rotations and the other people involved.
Nobody can return to a previous partner, so only a limited number of changes are possible within a small set of people.
Further changes must involve new labels and enlarge the supports, giving additional contributions
to $\Omega$.
If the people involved stop changing partners instead, their final degrees allow sharper
cost bounds.
We show that these additional contributions and sharper bounds cover the remaining costs, without using any contribution twice.
This proves \eqref{eq:summed-criterion-intro} and gives a rotation suitable for the branching induction.

\paragraph{Organization.}
In \Cref{sec:reduction} and~\Cref{sec:orders}, we reduce the problem of counting stable matchings to normalized instances and establish the required recurrence for downsets.
In \Cref{sec:weight}, we prove \Cref{thm:main}
assuming \Cref{thm:admissible-potential,thm:sum-admissible}.
In the same section, we show that \Cref{thm:admissible-potential} follows from the branching criterion \eqref{eq:summed-criterion-intro}.
We then prove this criterion: \Cref{sec:support-cost} reduces it to \Cref{prop:covering-budget}, which we prove in \Cref{sec:costs}--\Cref{sec:credit}.
Finally, in \Cref{sec:numbers}, we construct the weights required by \Cref{thm:sum-admissible}.

\section{Stable partners and normalized instances}
\label{sec:reduction}
Throughout this section, fix an instance $I$ with $n$ men and $n$ women.
For a matching $\mu$, write $\mu(v)$ for the woman matched to man $v$, and $\mu^{-1}(w)$ for the man matched to woman $w$.

We first recall facts about rotations and maximal chains that we will use.
We then construct a new instance $\widehat I$ by moving each person's stable partners to the beginning of their preference list.
We prove that $I$ and $\widehat I$ have the same stable partners and that every stable matching of $I$ is stable in~$\widehat I$.

\subsection{Rotations and maximal chains}
There is a stable matching in which every man receives his most preferred stable partner and every woman receives her least preferred stable partner.
This is the \emph{man-optimal stable matching}, denoted by $\mu^M$.
Similarly, the \emph{woman-optimal stable matching}, denoted by $\mu^W$, gives every woman her most preferred stable partner and every man his least preferred stable partner.
These properties follow from the optimality theorem of Gale and Shapley~\cite[Theorem~2]{GS};
see also the work of Gusfield~\cite[Section~2]{Gusfield1987}, and Cai and Thomas~\cite[Claim~2.1 and Corollary~2.3]{CT}.
These two matchings are the minimum and maximum elements, respectively, of the order defined below.

The stable matchings form a poset under the relation $\mu\le\nu$ when every man prefers his partner in $\mu$ to his partner in $\nu$, or has the same partner in both. With this convention, men's partners become worse and women's partners become better as we move upwards. 
A stable matching $\nu$ \emph{covers} $\mu$ if $\mu<\nu$ and there is no
stable matching $\eta$ with $\mu<\eta<\nu$.

For a stable matching $\mu$ and a man $v$, scan $v$'s preference list below $\mu(v)$ and find the first woman~$w$ who prefers $v$ to her partner under $\mu$.
Write $s_\mu(v)=w$; if there is no such woman, leave~$s_\mu(v)$ undefined.
We use the classic definition of a rotation by Gusfield~\cite[Section~3.1]{Gusfield1987}:

\begin{definition}
\label{def:rotation}
  An ordered list
  \begin{equation*}
    \rho=\big((v_0,w_0),\ldots,(v_{k-1},w_{k-1})\big),\qquad k\ge2,
  \end{equation*}
  of distinct pairs of $\mu$ is a \emph{rotation exposed in $\mu$} if
  $s_\mu(v_i)=w_{i+1}$ for every $i$, with indices modulo~$k$.
  Its \emph{elimination} replaces $v_iw_i$ by $v_iw_{i+1}$ and leaves all other pairs unchanged.
  The resulting matching is denoted $\mu/\rho$.
  The \emph{length} of $\rho$ is $k$.
\end{definition}

Eliminating a rotation makes a cyclic exchange of partners: each man $v_i$ receives $w_{i+1}$, with indices taken modulo $k$.
The resulting matching is stable \cite[Theorem~6]{Gusfield1987}.
Eliminating an exposed rotation gives exactly a cover in the order on stable matchings, as stated in the following:
\begin{lemma}[{\cite[Theorem~6]{Gusfield1987};
see also~\cite[Claim~4.2]{CT}}]
\label{lem:covers}
  A stable matching $\nu$ covers $\mu$ if and only if $\nu=\mu/\rho$ for a rotation $\rho$ exposed in $\mu$.
\end{lemma}

Let $\Pi=\Pi(I)$ be the set of all rotations exposed in at least one stable matching of $I$. In describing a rotation, the choice of starting pair is irrelevant.
In fact, starting the list at another pair and keeping the same cyclic order describes the same
rotation.
\begin{proposition}[{\cite[Theorems~2 and~6]{Gusfield1987};
see also~\cite[Claims~4.3(2) and~4.5]{CT}}]
\label{prop:rotation-chains}
  The following statements hold for the instance $I$.
  \begin{enumerate}
    \item[\rm(i)] Every maximal chain from $\mu^M$ to $\mu^W$ eliminates every rotation in $\Pi$ exactly once.
      Consequently, all such chains have $|\Pi|$ steps.
      Their orders of elimination may differ.
		\item[\rm(ii)] When a man participates in a rotation, his partner changes
		to the next woman in his list of stable partners, ordered
		from most to least preferred. When a woman participates,
		her partner changes to the next man in her list of stable
		partners, ordered from least to most preferred.
	\end{enumerate}
\end{proposition}

Fix a maximal chain of stable matchings of $I$:
\begin{equation*}
\label{eq:matching-chain}
\tag{MC}
  \mu_0=\mu^M<\mu_1<\cdots<\mu_R=\mu^W \enspace  \enspace .
\end{equation*}
Here, maximality means that no stable matching can be inserted between two consecutive matchings in the chain.
Thus, $\mu_i$ covers $\mu_{i-1}$ and is obtained from it by eliminating an exposed rotation, denoted by $\rho_i$.
By \Cref{prop:rotation-chains}, the list $\rho_1,\ldots,\rho_R$ contains every element of $\Pi$ exactly once, so $R=|\Pi|$.
If there is only one stable matching, then $\mu^M=\mu^W$, $\Pi$ is empty, and $R=0$.

Along \eqref{eq:matching-chain}, each man starts with his best stable partner and ends with his worst.
Each time his partner changes, he receives the next stable partner in his preference order.
He therefore receives every stable partner, without skipping any.
Each woman similarly receives all her stable partners, starting with her worst and ending with her best.

For a person $v$, let $D_v$ be the number of their stable partners and let $d_v$ be the number of rotations in $\Pi$ containing them.
Since every such rotation is eliminated exactly once along the chain, and each changes their partner,
\begin{equation}
\label{eq:degree-equals}
	d_v=D_v-1 \enspace .
\end{equation}

\subsection{Normalized instance}
Call an instance \emph{normalized} if each person ranks all their stable partners before all their other partners.
For a given instance $I$, construct $\widehat I$ by moving each person's stable partners
to the beginning of their preference list, preserving their relative order, and placing all remaining partners afterwards in their original relative order.

The next lemma constructs preference lists from a prescribed sequence of partner exchanges.
If no man or woman returns to a previous partner, we can choose the preferences so that each
exchange is a rotation elimination.
We will use this construction to obtain normalized instances.
\begin{lemma}
\label{lem:history-realization}
  Fix $n$ men and $n$ women, and let $\mu_0,\ldots,\mu_R$ be perfect matchings between them.
  Suppose each matching is obtained from the preceding one by cyclically exchanging the partners of at least two men, and no man or woman returns to a partner after leaving them.

  There exists a normalized instance $I$ in which $\mu_0,\ldots,\mu_R$ is a maximal chain of stable matchings, with each prescribed exchange eliminating an exposed rotation.
  The stable pairs of $I$ are exactly the pairs appearing in these matchings.
\end{lemma}
\begin{proof}
  For each man, rank his partners in the order in which he receives them in $\mu_0,\ldots,\mu_R$, from most to least preferred.
  For each woman, rank her partners in the reverse order, so that she prefers each new partner to her previous one.
  For every person, place all partners not encountered in the sequence after the encountered partners, in any order.
  These preference lists define the instance $I$.

  Consider a matching $\mu_i$ in the sequence.
  If a man prefers a woman to $\mu_i$'s partner, she was his partner earlier in the 	sequence.
  Since their partnership ended, she has received a later partner, whom she prefers to him. Thus, no pair blocks $\mu_i$.
	
  At each step, a participating man's next partner is immediately below his current partner on his list. She prefers him to the man she leaves.
  Hence the specified exchange is an exposed rotation, and the step is a cover by \Cref{lem:covers}.
  The first matching gives every man his first choice, and the last gives every woman her first choice.
  The sequence is therefore a maximal chain between the two optimal matchings.
  By \Cref{prop:rotation-chains}, a maximal chain visits every stable partner, so the encountered pairs are exactly the stable pairs of this instance.
  This also proves normalization.
\end{proof}

\begin{proposition}
\label{prop:normalization}
  Every stable matching of $I$ is stable in $\widehat I$.
  The two instances have the same stable partners for every person, and every maximal chain of $I$ remains a maximal chain of $\widehat I$ with the same rotations.
  In particular, $\widehat I$ is normalized and $s(I)\le s(\widehat I)$.
\end{proposition}
\begin{proof}
  Apply \Cref{lem:history-realization} to a fixed maximal chain of $I$, keeping the original order of partners not encountered on the chain.
  Each person encounters all their stable partners, in preference order for men and reverse preference order for women.
  Thus the resulting lists are exactly those of $\widehat I$.
  The lemma proves that $\widehat I$ is normalized, has the same stable partners as $I$, and retains the chain with the same rotations.
  The same argument applies to every maximal chain of $I$.
	
  Every stable matching of $I$ belongs to a maximal chain of $I$.
  Since that chain remains a maximal chain of $\widehat I$, every stable matching of $I$ is stable in $\widehat I$.
  Hence $s(I)\le s(\widehat I)$.
\end{proof}

Since every stable matching of $I$ is also stable in $\widehat I$, it suffices to bound the number of stable matchings of normalized instances.
Moreover, $I$ and $\widehat I$ have the same stable partners for every person. Thus replacing $I$ by $\widehat I$ does not change any of the numbers $D_v$ appearing in
\Cref{thm:main}.

\section{The rotation poset and downsets}
\label{sec:orders}
In this section, given an instance $I$, we use the fixed maximal chain~\eqref{eq:matching-chain} to define a partial order on $\Pi=\Pi(I)$.
When $I$ is normalized, we show that the downsets of this poset are in bijection with the stable matchings of $I$.
We also show that deleting a downset or an upset of rotations leaves the rotation poset of another normalized instance, as needed for the induction.

Let $\rho_1,\ldots,\rho_R$ be the rotations of $I$, listed in the order in which they are eliminated along the fixed maximal chain~\eqref{eq:matching-chain}.
Define a relation $\preceq$ on $\Pi$ by writing $\rho_i\preceq\rho_j$ if $i=j$, or if there are indices
\begin{equation*}
  i=i_0<i_1<\cdots<i_t=j
\end{equation*}
such that $\rho_{i_{\ell-1}}$ and $\rho_{i_\ell}$ contain a common man for all $\ell = 1,\hdots,t$.
This relation is a partial order: it is reflexive and transitive by construction, and antisymmetric because every comparison between distinct rotations follows the increasing indices.
We call it the \emph{shared-participant order}.
Let $Q=(\Pi,\preceq)$ be the resulting poset. 
For each man $v$, the rotations containing $v$ form a chain in~$Q$, which we denote by $C_v$. 
The rotations in~$C_v$ are ordered by their elimination along~\eqref{eq:matching-chain}.
The chains $C_v$ may overlap, because a rotation belongs to the chain of every man participating in it.
All comparisons in $Q$ follow from these chains by transitivity.
For example, if $\rho$ precedes $\sigma$ on~$C_u$, and~$\sigma$ precedes $\tau$ on $C_v$, then $\rho\preceq\tau$, even if $\rho$ and $\tau$ have no man in common.
More generally, $\rho\preceq\tau$ for distinct rotations exactly when one can go from $\rho$ to $\tau$ through a sequence of rotations, with each consecutive pair appearing in that order on some man's chain.
In this sense, the chains~$C_v$ generate~$Q$.

For each rotation $\rho_i\in\Pi$, its participating men and their partners before and after elimination are specified by $\mu_{i-1}$ and $\mu_i$, respectively.
We label each woman by the man matched to her in~$\mu_0$.

An \emph{upset} of $Q$ is a subset $U\subseteq\Pi$ such that $\rho\in U$ and $\rho\preceq\tau$ imply $\tau\in U$.
For each $\rho\in\Pi$, put
\begin{equation*}
  \past\rho=\{\tau\in\Pi:\tau\preceq\rho\},
  \qquad
  \future\rho=\{\tau\in\Pi:\rho\preceq\tau\} \enspace .
\end{equation*}
Thus $\past\rho$ contains $\rho$ and all rotations smaller than it, while $\future\rho$ contains $\rho$ and all rotations larger than it.
The \emph{proper past} of $\rho$ is $\past\rho\setminus\{\rho\}$, consisting of all rotations
$\tau\prec\rho$.

Whenever we delete elements from $Q$, we retain the same comparisons between the remaining elements; that is, we take the induced poset.
A \emph{linear extension} of $Q$ is a list of all its elements, each appearing exactly once, in which $\rho$ appears before $\tau$ whenever $\rho\preceq\tau$ and $\rho\ne\tau$.

\paragraph{Remark.}
Karlin, Oveis Gharan, and Weber~\cite{KGW} and Palmer and P\'alv\"olgyi~\cite{PP} formulate their bounds using the \emph{full rotation order}, denoted here by $\prec_{\mathrm{f}}$.
For distinct rotations $\rho$ and $\sigma$, we write $\rho\prec_{\mathrm{f}}\sigma$ if every rotation elimination sequence starting from $\mu^M$ and containing $\sigma$ eliminates $\rho$ before $\sigma$.
We instead work with the shared-participant order $\preceq$, generated by the order of rotations along each man's chain.
This description lets us track partner changes and construct the smaller instances needed for induction.
For normalized instances, however, the two orders agree.
We prove this below and then use the description through participant chains.

\medskip
 
Let $Q_{\rm f}=(\Pi,\prec_{\rm f})$ be the full rotation poset of $I$.
\begin{lemma}[Irving--Leather~\cite{IL}]
\label{lem:full-correspondence}
  Let $I$ be a stable matching instance, and let $Q_{\rm f}$ be its full rotation poset. For every downset $D$ of $Q_{\rm f}$, eliminating the rotations in $D$ from $\mu^M$, in any order
consistent with $Q_{\rm f}$, gives a stable matching $\mu_D$.
  The map $D\mapsto\mu_D$ is a bijection between the downsets of $Q_{\rm f}$ and the stable matchings of $I$.
  In particular, $s(I)=J(Q_{\rm f})$.
\end{lemma}

\subsection{Reordering rotations in one instance}
We first show that, for normalized instances, the two orders $\preceq$ and $\prec_{\rm f}$  agree.

\begin{lemma}
\label{lem:orders-coincide}
  If $I$ is normalized, then $Q=Q_{\rm f}$.
\end{lemma}
\begin{proof}
  By \Cref{prop:rotation-chains}, rotations involving the same man must follow his sequence of stable partners.
  Thus every comparison in $Q$ also holds in $Q_{\rm f}$.

  Conversely, take any linear extension of $Q$.
  It can be obtained from $\rho_1,\ldots,\rho_R$ by swapping adjacent incomparable rotations.
  At each swap, the two exchanges involve disjoint sets of men. They also involve disjoint sets of women: if a woman participated in both exchanges, her partner between them would be a man participating in both. Therefore the exchanges can be performed in either order, with the same matching after both.

  These swaps preserve each man's and woman's sequence of partners.
  By \Cref{prop:rotation-chains}, the original chain visits every stable partner, in preference order for men and reverse preference order for women.
  Since $I$ is normalized, the construction in \Cref{lem:history-realization}, applied
to the reordered history and keeping the original order of all remaining partners, gives exactly $I$.
  Hence every linear extension of~$Q$ is a valid elimination order in $I$.
  It follows that $Q_{\rm f}$ has no additional comparisons: otherwise, some linear extension of $Q$ would reverse one of them.
  Thus $Q=Q_{\rm f}$.
\end{proof}

\begin{lemma}
\label{lem:reorder}
  Suppose that $I$ is normalized.
  \begin{enumerate}[label=\textup{(\roman*)},ref=\thelemma\,(\roman*)]
    \item Every linear extension $\tau_1,\ldots,\tau_R$ of $Q$ gives a valid sequence of rotation eliminations from $\mu_0$ to $\mu_R$. Each man and woman receives the same successive partners, in the same order, as along~\eqref{eq:matching-chain}.
    \item\label[lemma]{lem:reorder-iii}
    For every downset $D$ of $Q$, there is such a sequence eliminating all rotations in $D$ before those outside $D$.
  \end{enumerate}
\end{lemma}
\begin{proof}
  By \Cref{lem:orders-coincide}, $Q=Q_{\rm f}$.
  Thus every linear extension gives a valid elimination sequence by \Cref{lem:full-correspondence}.
  The assertion about successive partners follows from \Cref{prop:rotation-chains}.

  For (ii), concatenate a linear extension of $D$ with a linear extension of $Q\setminus D$. Since $D$ is a downset, this is a linear extension of $Q$, so (i) applies.
\end{proof}

Recall that each woman is labelled by her partner in $\mu_0$.
The next lemma shows that every downset involves the same labels among men and women.

\begin{lemma}
\label{lem:twopaths}
  Suppose that $I$ is normalized.
  \begin{enumerate}[label=\textup{(\roman*)},ref=\thelemma\,(\roman*)]
    \item For every pair matched in $\mu_0$, either neither partner participates in any rotation, or both have the same first rotation.
    \item For every downset $D$ of $Q$, the set of labels of the men participating in rotations of $D$ equals the set of labels of the women participating in rotations of $D$.
  \end{enumerate}
\end{lemma}
\begin{proof}
  For (i), let $v$ and $w$ be matched in $\mu_0$.
  They remain matched until the first rotation involving either of them.
  That rotation must involve both, so their first rotations coincide.
  If neither participates, they remain matched throughout.

  For (ii), since $Q=Q_{\rm f}$, the rotations involving any fixed man or woman form a chain in $Q$.
  Thus, a man or woman participates in a rotation of $D$ if and only if their first rotation belongs to $D$.

  By (i), the two partners in each initial pair therefore either both participate in rotations of $D$ or neither does.
  Since they have the same label, the two sets of labels are equal.
\end{proof}

\begin{proposition}
\label{prop:reduction}
  If $I$ is normalized, then $s(I)=J(Q(I))$.
  For an arbitrary instance $I$, let $\widehat I$ be its normalization.
  Then $s(I)\le s(\widehat I)=J(Q(\widehat I))$.
\end{proposition}
\begin{proof}
  If $I$ is normalized, \Cref{lem:orders-coincide} gives $Q=Q_{\rm f}$, so the first assertion follows from \Cref{lem:full-correspondence}.
  Applying this to $\widehat I$ and using \Cref{prop:normalization} proves the second assertion.
\end{proof}

\subsection{Instances obtained by deleting downsets and upsets}
We now prescribe that certain rotations have been eliminated and that certain others will not be eliminated. The next lemma realizes the remaining matchings as the stable matchings of another normalized instance.
This ensures that the induction stays within the class of posets just described.

\begin{lemma}
\label{lem:residual}
  Let $Q$ be the rotation poset of a normalized instance $I$.
  Let $D$ be a downset and $U$ an upset of $Q$, with $D\cap U=\varnothing$, and put $P=Q\setminus(D\cup U)$.
  There is a normalized instance $I_P$ on the same people whose rotation poset is the induced poset on $P$.
  Each person's rotation chain consists of their retained rotations.
  The map
  \begin{equation*}
	K\longmapsto D\cup K
  \end{equation*}
  is a bijection from the downsets of $P$ to the downsets of $Q$ that contain $D$ and avoid $U$. Corresponding downsets give the same matching in $I_P$ and $I$.
\end{lemma}
\begin{proof}
  Choose a linear extension with $D$ first, $P$ next, and $U$ last.
  This is possible because $D$ is a downset and $U$ is an upset.
  By \Cref{lem:reorder}, it gives an elimination sequence with the same successive partners as the original chain.
  Retain the portion just after $D$ and just before $U$.
  Apply \Cref{lem:history-realization} to this portion.
  It gives a normalized instance $I_P$ with precisely the retained rotations and partner changes.
	
  The set $P$ is order-convex: if $a\preceq b\preceq c$ and $a,c\in P$, then $b\in D$ would imply $a\in D$, while $b\in U$ would imply $c\in U$.
  Thus $b\in P$.
  Consequently, every path of common-participant comparisons between retained rotations stays in $P$.
 The retained chains therefore generate exactly the induced order on $P$.
  By definition, this is the shared-participant order of $I_P$.
  Since $I_P$ is normalized, \Cref{lem:orders-coincide} identifies it
  with the full rotation order of $I_P$.
  Each person's degree is the number of their retained rotations.
	
  For a downset $K$ of $P$, every predecessor in $Q$ of an element of $K$ lies in $D\cup K$; it cannot lie in~$U$, since $U$ is an upset.
  Hence $D\cup K$ is a downset of $Q$ containing $D$ and avoiding $U$. Conversely, deleting $D$ from any such downset gives a downset of $P$.
  These maps are inverse.
  The initial matching of $I_P$ is the matching reached after $D$, and the retained rotations make the same partner changes.
  Thus, corresponding downsets give the same matching.
\end{proof}

\begin{lemma}
\label{lem:recursion}
  For every rotation $\rho\in Q$,
  \begin{equation}
  \label{eq:recursion}
   J(Q)=J(Q\setminus\past\rho)+J(Q\setminus\future\rho) \enspace .
  \end{equation}
  Both smaller posets are rotation posets of normalized instances on the same people, with degrees equal to their numbers of retained occurrences.
\end{lemma}
\begin{proof}
  The recurrence holds for every finite poset: partition its downsets according to whether they contain $\rho$; see the work of Karlin, Oveis Gharan and Weber~\cite[Section~3]{KGW}.
  Apply \Cref{lem:residual} with $(D,U)=(\past\rho,\varnothing)$ or $(D,U)=(\varnothing,\future\rho)$, respectively, to obtain normalized instances whose rotation posets are $Q\setminus\past\rho$ and $Q\setminus\future\rho$.
\end{proof}

For the rest of the proof, $Q$ is the rotation poset of a normalized instance, together with a fixed maximal chain and its cyclic partner changes.
This convention includes every instance obtained by \Cref{lem:residual}.
Label each woman by her partner in the initial matching of the current chain.
Write $(v_1,\ldots,v_k)$ for the rotation in which $v_i$ receives the current partner of $v_{i+1}$, with indices modulo $k$; when $k=2$, this is a swap.
The number of rotations containing a person is their \emph{degree}.
The word ``new'' means that a person has not yet occurred in the current chain.
Each participation of a person in a rotation is called an ``occurrence'' of that person.
Formally, an \emph{occurrence} is a pair $(v,\rho)$, where person $v$ participates in rotation $\rho$.
Its \emph{rank} is the position of $\rho$ on that person's rotation chain.
For example, if $v$ participates in rotations $\rho_1,\rho_2,\rho_3$, in that order, then $v$ has three occurrences, of ranks $1,2,3$, respectively.
Two people participating in the same rotation give two distinct occurrences.

\section{Building the branching potential}
\label{sec:weight}
In this section, we define a function $F_h$ on the family $\mathcal Q$ of rotation posets of normalized instances on fixed sets of $n$ men and $n$ women.
For each $Q\in\mathcal Q$, we assign each man and woman a weight depending on their degree.
The weights are determined by a fixed sequence $(h_d)_{d\ge1}$.
To ensure that $F_h$ is a branching potential, we impose certain conditions on $(h_d)$.
If its sum $S=\sum_{d\ge1}h_d$ is finite, then $F_h(Q)\le nS$, giving an upper bound of $e^{nS}$ on the number of stable matchings.
Thus we seek a sequence with a small sum that also satisfies the conditions needed for the branching argument.
We collect these conditions below and call sequences satisfying them \emph{admissible}.
Their role will become clear as the proof proceeds.
In \Cref{sec:numbers}, we construct an admissible sequence with $S<\ln2.81$, completing the proof of \Cref{thm:main}.

\subsection{Admissible sequences}
Let $h_1\ge h_2\ge\cdots>0$.
Set $W(0)=H_0=0$, and
\begin{equation*}
  W(d)=\sum_{r=1}^d h_r\quad\text{and}\quad
  H_d=\sum_{r=1}^d(r+1)h_r \enspace. 
\end{equation*}
Also, for $t>0$, define
\begin{equation*}
  \lambda(t) = -\ln(1-e^{-t}) \enspace .
\end{equation*}

\begin{definition}
\label{def:admissible}
  A sequence $h_1\ge h_2\ge\cdots>0$ is \emph{admissible} if it satisfies the following five conditions, with $A_1=4h_1$ and $A_c=H_c+h_1+2h_2$ for $c\ge2$:
  \begin{enumerate}[label=\textup{(H\arabic*)},ref=H\arabic*,leftmargin=3em,itemsep=5pt]
    \item\label{cond:first} $2h_1\ge\ln2$.
    \item\label{cond:degree} $\lambda(2h_d)\le H_d+h_2$ for every $d\ge2$.
    \item\label{cond:exceptional} $\lambda(2h_a)+\lambda(h_{a-1}+2h_a+h_c) \le H_a+A_c$ for every $a\ge3$ and $c\ge1$.
    \item\label{cond:transfer} $h_1\ge2h_2$, $2h_3\ge h_2$, and $4h_1-\lambda(h_1)\ge h_2$.
    \item\label{cond:stop} $h_3\le1/10$ and $h_2\le8h_3/5$.
  \end{enumerate}
\end{definition}

Fix sets of $n$ men and $n$ women.
Let $\mathcal Q$ be the family of posets obtained from normalized instances on these people, each with its specified rotations and a chosen maximal chain.
By \Cref{lem:residual}, deleting a principal downset or a principal upset gives another member of $\mathcal Q$.
Fix an arbitrary admissible sequence $(h_r)_{r\ge1}$, and for every $Q\in\mathcal Q$ define
\begin{equation*}
  F_h(Q) = \frac12\sum_{v=1}^n \bigl(W(d_v^m)+W(d_v^w)\bigr) \enspace .
\end{equation*}
As we proceed with the proof, we explain how conditions \textup{(H1)}--\textup{(H5)} ensure that $F_h$ is a branching potential.

We want the potential to be small, but to decrease enough when rotations are deleted. If the weights have finite sum
$S=\sum_{r\ge1}h_r$, then $W(d)\le S$ and $F_h(Q)\le nS$.
When a person's degree falls from $d$ to $d-1$, the value $W(d)$ decreases by $h_d$. 
Since  $(h_d)$ is decreasing, this loss is larger for smaller degrees.

Let $p=F_h(Q)-F_h(Q\setminus\past\rho)$.
To ensure $F_h$ is branching potential, we need, for some $\rho$,$F_h(Q)-F_h(Q\setminus\future\rho)\ge\lambda(p)$ (see \eqref{eq:good} below).
For small $p$, the value $\lambda(p)$ is approximately $\ln(1/p)$.
Thus, if deleting $\past\rho$ causes only a small decrease in $F_h$, deleting $\future\rho$ must cause a large decrease.
The fact that partners never repeat helps us obtain the required decreases.
Consider men and women separately.
For each person who participates in a rotation, delete their first rotation and all its successors, and sum the resulting decreases in $\sum_v W(d_v)$.
We will show that a person's increment~$h_r$ is counted at least $r+1$ times in this sum, because by their $r^{\textnormal{th}}$ rotation they have had $r+1$ distinct partners;
see \Cref{lem:upset-sum}.

Thus a person of degree $d$ contributes at least $H_d=\sum_{r=1}^d(r+1)h_r$ to the sum.
We want the weights to have a finite sum, while $H_d$ grows fast enough to satisfy \eqref{cond:degree}.
This suggests choosing $h_d\sim C/d^2$: then $\sum_{d\ge1}h_d$ is finite, whereas
\begin{equation*}
  H_d\sim C\ln d
  \quad\text{and}\quad
  \lambda(2h_d)=2\ln d+O(1) \enspace .
\end{equation*}
Taking $C>2$ therefore makes $H_d$ grow faster than $\lambda(2h_d)$.
In \Cref{sec:numbers}, we use
\begin{equation*}
  h_d=\frac{C}{d(d-1)}\qquad\text{for }d\ge33,
\end{equation*}
which satisfies $h_d\sim C/d^2$ and has the exact sum $\sum_{d=33}^{\infty}h_d=C/32$.
We choose $C>2$ and the first $32$ weights by finite convex optimization, ensuring that all admissibility conditions hold and that $S<\ln2.81$.
Then
\begin{equation*}
  H_d=C\ln d+O(1),
  \qquad
  \lambda(2h_d)=2\ln d+O(1) \enspace .
\end{equation*}
Hence~$H_d$ eventually grows faster than $\lambda(2h_d)$,
as needed for \eqref{cond:degree}.
Moreover, $\sum_{d=33}^{\infty}h_d=C/32$, so the total weight is finite and easy to calculate.

We also record some properties of $\lambda$ used below.
It is decreasing and convex, since \mbox{$\lambda'(t)=-1/(e^t-1)<0$} and $\lambda''(t)=e^t/(e^t-1)^2>0$.
We will use
\begin{equation*}
  \lambda(t)-\lambda(2t)=\ln(1+e^{-t})\le\ln2
\end{equation*}
and
\begin{equation}
\label{eq:lambda-difference}
 \lambda(y)-\lambda(x)\le\ln(x/y)\quad\text{for}~0<y\le x \enspace .
\end{equation}
The latter follows by integrating $-\lambda'(t)=1/(e^t-1)\le1/t$ from $y$ to $x$.

\subsection{Labels and branch losses}
For each $Q\in\mathcal Q$, let $\mu_Q$ be the man-optimal matching of its associated instance. This is the initial matching of its chosen chain.
Label each woman by her partner in $\mu_Q$; thus the woman with label $v$ is $\mu_Q(v)$. In an argument concerning men, the person with label $v$ means man $v$; in an argument concerning women, it means woman $\mu_Q(v)$.
These labels remain fixed as partners change.
When we pass to a smaller poset $P$, we use its own initial matching $\mu_P$. First rotations, ranks, and women's labels are then defined anew.

Arguments that apply separately to men and women are written for men; the same arguments apply to women after interchanging the sexes.
In these arguments, we omit $m$ and $w$ from the notation.
For example, $d_v$ denotes the degree of the person with label $v$, with the sex clear from the context.
When we need to distinguish men from women, we write $d_v^m$ and $d_v^w$.
We use the same convention for the other quantities below.

For one sex, put
\begin{equation*}
 F(Q)=\sum_{v=1}^n W(d_v) \enspace .
\end{equation*}
Write $F_m(Q)=\sum_{v=1}^n W(d_v^m)$ and $F_w(Q)=\sum_{v=1}^n W(d_v^w)$.
Then the potential defined above satisfies
\begin{equation*}
  F_h(Q)=\frac{F_m(Q)+F_w(Q)}2 \enspace .
\end{equation*}
All these potentials are nonnegative and vanish on an empty poset.
In a smaller instance, degrees count the retained occurrences.
Relabelling women does not change the sums, which depend only on the degrees of the same people.
For a rotation $\rho$, define the one-sex losses
\begin{equation*}
  p(\rho)=F(Q)-F(Q\setminus\past\rho),\qquad q(\rho)=F(Q)-F(Q\setminus\future\rho) \enspace .
\end{equation*}
The losses for $F_h$ are their averages:
\begin{equation*}
  p^*(\rho) = \frac{p_m(\rho)+p_w(\rho)}2,\qquad q^*(\rho)=\frac{q_m(\rho)+q_w(\rho)}2 \enspace .
\end{equation*}
Every loss is positive, since the deleted set contains $\rho$.
By \Cref{lem:recursion}, to prove that $F_h$ is a branching potential, it suffices to find, for every nonempty $Q\in\mathcal Q$, a rotation $\rho\in Q$ such that
\begin{equation}
\label{eq:good}
  e^{-p^*(\rho)}+e^{-q^*(\rho)}\le1 \enspace .
\end{equation}
Rearranging and taking logarithm of both sides gives the equivalent inequality $\lambda(p^*(\rho))\le q^*(\rho)$.



\paragraph{Active labels and first rotation $b_v$.} Call a label $v$ \emph{active} if $d_v^m>0$, and put
\begin{equation*}
  V=\{v:d_v^m>0\}=\{v:d_v^w>0\} \enspace .
\end{equation*}
The equality follows from \Cref{lem:twopaths}.
The man and woman with active label $v$ have the same first rotation, denoted by $b_v$. All these choices refer to the current poset $Q$.
Unless an explicit range is given, subsequent sums over labels are over active labels $V$.
A degree-zero person has no first rotation and contributes zero to $F_h$ and to the sums of $H_{d_v}$.
Labels sharing the same first rotation are counted separately.

\begin{proposition}
\label{prop:summed-criterion}
  For every admissible sequence and every nonempty $Q\in\mathcal Q$,
  \begin{equation*}
  \label{eq:summed-criterion}
  \tag{BC}
    \sum_{v\in V}\lambda(p^*(b_v)) \le\sum_{v\in V}q^*(b_v) \enspace . 
  \end{equation*}
\end{proposition}
 We prove this proposition in \Cref{sec:BC-proof}.
Before doing so, we show that it implies that $F_h$ is a branching
potential, and then use this to prove the main theorem.
\begin{proof}[\bfseries Proof of \Cref{thm:admissible-potential} assuming \Cref{prop:summed-criterion}]
  Fix an admissible sequence $h$ and a nonempty $Q\in\mathcal Q$.
  There is at least one active label.
  By \Cref{prop:summed-criterion}, some $v\in V$ satisfies
  \begin{equation*}
    \lambda(p^*(b_v))\le q^*(b_v) \enspace .
  \end{equation*}
  Thus $\rho=b_v$ satisfies \eqref{eq:good}.
  The downsets containing and avoiding $\rho$ are counted by $Q\setminus\past\rho$ and $Q\setminus\future\rho$, respectively, by \Cref{lem:recursion}.
  Both residuals belong to $\mathcal Q$ by \Cref{lem:residual}, and both have fewer rotations.
  Together with $F_h(\varnothing)=0$, these facts show that $F_h$ is a branching potential.
\end{proof}

\begin{proof}[\bfseries Proof of \Cref{thm:main} assuming \Cref{thm:admissible-potential,thm:sum-admissible}]
  By \Cref{thm:sum-admissible}, choose an admissible sequence $(h_r)_{r\ge1}$ such that
	$S:=\sum_{r\ge1}h_r<\ln 2.81$.
  Let $\widehat I$ be the normalized instance associated with~$I$, and let $\widehat Q$ be its rotation poset.
  By \Cref{prop:reduction}, \Cref{thm:admissible-potential}, and \eqref{eq:branching-potential},
  \begin{equation*}
	s(I) \le s(\widehat I)
         = J(\widehat Q)
         \le e^{F_h(\widehat Q)} \enspace .
  \end{equation*}
  Normalization preserves every person's stable partners, so \eqref{eq:degree-equals} gives
  \begin{equation*}
	F_h(\widehat Q)=\frac12\sum_{v=1}^n \bigl(W(D_v^m-1)+W(D_v^w-1)\bigr) \enspace .
  \end{equation*}
  This proves the degree bound in \Cref{thm:main}.
  Since every degree is at most $n-1$,
  \begin{equation*}
    F_h(\widehat Q)\le nW(n-1)\le nS<n\ln 2.81 \enspace .
  \end{equation*}
  Thus, $s(I)<2.81^n$.
  Since $I$ was arbitrary, $\SM(n)<2.81^n$.
\end{proof}


\section{Supports and costs}
\label{sec:support-cost}
We now develop the estimates needed to prove \eqref{eq:summed-criterion}.
We first calculate the sum of the losses $q(b_v)$ by counting how often each weight $h_r$ appears.
We then group labels with the same first rotation and estimate the corresponding terms $\lambda(p(b_v))$.
These estimates apply separately to men and women.
They handle some groups directly and leave the remaining groups for the later sections.

\subsection{Supports and deletion losses}

For a rotation $\rho$, define its \emph{support} by
\begin{equation}
\label{eq:support}
  S(\rho)=\{v\in V:b_v\preceq\rho\} \enspace .
\end{equation}
Write $s(\rho)=|S(\rho)|$.
These are exactly the  labels participating in $\past\rho$.
Indeed, a person's first rotation precedes all their other rotations.
The support is the same for men and women by \Cref{lem:twopaths}, and supports increase along
the partial order.

For one sex, list a person's rotations as $\rho_{v,1},\ldots,\rho_{v,d_v}$ in their order on that person's chain.
Thus,~$\rho_{v,r}$ is an occurrence of rank $r$.

\begin{lemma}
\label{lem:support}
  Let $\rho_{v,r}$ be the $r^{\textnormal{th}}$ rotation containing person $v$.
  Then
  \begin{equation}
  \label{eq:support-lower}
	s(\rho_{v,r})\ge r+1 \enspace .
  \end{equation}
\end{lemma}
\begin{proof}
  We first consider a man $v$ and put $\rho=\rho_{v,r}$.
  By \Cref{lem:reorder-iii}, we may eliminate all rotations in $\past\rho$ before eliminating any other rotation.
  Every man participating in these rotations has a label in $S(\rho)$. Initially, the men with these labels are matched to the women with the same labels.
  Each rotation only redistributes their current partners among its participants.
  Thus, these men remain matched to exactly the women with labels in $S(\rho)$ throughout the elimination.
  Exactly $r$ rotations in $\past\rho$ contain $v$: they are his first $r$ rotations.
  Each gives him a new partner, and he never returns to an earlier partner.
  Including his initial partner, he therefore has $r+1$ distinct partners.
  All their labels belong to $S(\rho)$, so $s(\rho)=|S(\rho)|\ge r+1$.
  The same argument applies to women: by \Cref{lem:twopaths}, $S(\rho)$ is also the set of labels of women participating in $\past\rho$.
\end{proof}

Define the \emph{support excess} by
\begin{equation}
\label{eq:omega}
  \Omega=\sum_{v\in V}\sum_{r=1}^{d_v} \bigl(s(\rho_{v,r})-r-1\bigr)h_r \enspace .
\end{equation}
Every summand is nonnegative.
We call the summand indexed by $(v,r)$ that occurrence's contribution to $\Omega$.

\begin{lemma}
\label{lem:upset-sum}
  For either sex,
  \begin{equation}
  \label{eq:upsetsum}
    \sum_{v\in V} q(b_v)=\sum_{v\in V} H_{d_v}+\Omega \enspace .
  \end{equation}
  The first occurrence of person $v$ contributes $(s(b_v)-2)h_1$ to $\Omega$.
\end{lemma}
\begin{proof}
  Fix one sex and a person $u\in V$, whose rotations are $\rho_{u,1},\ldots,\rho_{u,d_u}$.
  For each $v\in V$, the upset~$\future b_v$ intersects this chain in a suffix, possibly empty.
  Deleting this suffix decreases $u$'s contribution by the sum of the corresponding weights $h_r$.
  Hence,
  \begin{equation*}
    q(b_v)=\sum_{u\in V}\sum_{r=1}^{d_u} h_r\mathbf 1_{\{b_v\preceq\rho_{u,r}\}} \enspace .
  \end{equation*}
  Summing over $v$ and interchanging the sums gives
  \begin{equation*}
    \sum_{v\in V}q(b_v)=\sum_{u\in V}\sum_{r=1}^{d_u} h_r\sum_{v\in V}\mathbf 1_{\{b_v\preceq\rho_{u,r}\}}=\sum_{u\in V}\sum_{r=1}^{d_u}s(\rho_{u,r})h_r=\sum_{u\in V}H_{d_u}+\Omega \enspace .
  \end{equation*}
  The inner sum counts the labels in $S(\rho_{u,r})$.
  The last equality follows from $H_d=\sum_{r=1}^d(r+1)h_r$ and~\eqref{eq:omega}.
  Finally, $\rho_{u,1}=b_u$, so the first occurrence of $u$ contributes $(s(b_u)-2)h_1$ to $\Omega$.
\end{proof}

\subsection{First groups}
For a rotation $\rho$ that is first for at least one person, put
\begin{equation*}
  V_\rho=\{v\in V:b_v=\rho\} \enspace .
\end{equation*}
We call $V_\rho$ a \emph{first group}. 
Since $b_v=\rho$ for every $v\in V_\rho$, we have $V_\rho\subseteq S(\rho)$.
The first groups partition~$V$.

\begin{lemma}
\label{lem:first-group-cost}
  Let $V_\rho$ be a first group and put $k=|V_\rho|$.
  Then, for either sex,
  \begin{equation}
  \label{eq:first-past}
    p(\rho)\ge\sum_{v\in V_\rho}h_{d_v},
  \end{equation}
  and
  \begin{equation}
  \label{eq:group-convexity}
    k\lambda(p(\rho)) \le\sum_{v\in V_\rho}\lambda(kh_{d_v}) \enspace .
  \end{equation}
\end{lemma}
\begin{proof}
  For each $v\in V_\rho$, the only rotation in $\past\rho$ containing $v$ is $\rho$ itself, since it is $v$'s first rotation.
  Deleting $\past\rho$ therefore changes the term $W(d_v)$ to $W(d_v-1)$, with $W(d_v)-W(d_v-1)=h_{d_v}$.
  All other terms in $F(Q)$ cannot increase under deletion.
  Summing these up  gives \eqref{eq:first-past}.	
  Since $\lambda$ is decreasing and convex,
  \begin{equation*}
	\lambda(p(\rho))
	\le\lambda\left(\sum_{v\in V_\rho}h_{d_v}\right)
	=\lambda\left(\frac1k\sum_{v\in V_\rho}kh_{d_v}\right)
	\le\frac1k\sum_{v\in V_\rho}\lambda(kh_{d_v}) \enspace .
  \end{equation*}
  Multiplying by $k$ gives \eqref{eq:group-convexity}.
\end{proof}

We next consider first groups of size one.
Write $V_\rho=\{x\}$, so $\rho$ is the first rotation of $x$.
Since every rotation has at least two participants, $\rho$ also contains someone who has participated earlier.
That earlier rotation involves at least two labels and does not involve $x$.
Hence $s(\rho)\ge3$.

When $s(\rho)=3$, the first occurrence of $x$ contributes only $h_1$ to $\Omega$, compared with at least $2h_1$ when $s(\rho)\ge4$.
This smaller contribution is why groups $V_\rho$ with $s(\rho)=3$ need separate treatment.
The next lemma describes their structure exactly.

\begin{lemma}
\label{lem:three-labels}
  Suppose $V_\rho=\{x\}$ and $s(\rho)=3$.
  Then $\past\rho=\{A,\rho\}$, where $A$ is necessarily a swap, say $A=(u,v)$, and is the first rotation of both $u$ and $v$.
  Moreover, $\rho=(u,x)$ or $\rho=(v,x)$.
  Each of $u,v$ can take part in at most one such swap with a new person, which must be their second rotation.
\end{lemma}
\begin{proof}
  We describe the rotations using men.
  Write $S(\rho)=\{u,v,x\}$.
  Since $\rho$ is the first rotation of~$x$, no rotation strictly before $\rho$ contains $x$.
  Every such rotation must therefore swap $u$ and $v$.	
  There is at least one earlier rotation, since $x$ is the only label whose first rotation is $\rho$.
  There cannot be two, since both would swap $u$ and $v$.
  The second swap would return them to their initial partners, which is impossible.
  Thus, $\past\rho=\{A,\rho\}$,  $A=(u,v),$ and $A$ is the first rotation of both $u$ and $v$.
	
  After eliminating $A$, the partners of $u,v,x$ have labels $v,u,x$, respectively.
  If $\rho$ involved all three men, one of $u,v$ would receive his initial partner again.
  Hence, $\rho$ must be a swap: either $(u,x)$ or~$(v,x)$.
  This swap is the second rotation of the participant other than $x$.
  Since each person has only one second rotation, each of $u,v$ can participate in at most one such swap with a person moving for the first time.
  The same argument applies with the sexes interchanged.
\end{proof}

\subsection{Costs of first groups}
\label{sec:account}
We call $\lambda(p^*(b_v))$ the \emph{cost} of label $v$.
To prove \eqref{eq:summed-criterion}, we group the costs on its left-hand side according to the first groups.
All labels in $V_\rho$ have first rotation $\rho$, so their total cost is $|V_\rho|\lambda(p^*(\rho))$.
We begin with one sex and omit sex indices.
The corresponding cost of $V_\rho$ is $|V_\rho|\lambda(p(\rho))$.
For the corresponding one-sex sum on the right-hand side of \eqref{eq:summed-criterion}, \Cref{lem:upset-sum} gives $\sum_v q(b_v)=\sum_v H_{d_v}+\Omega$.
To relate the costs to the right-hand side, we will compare them with the degree terms
$H_{d_v}$ and the support excess $\Omega$.
Every statement in this section applies separately to each sex.

\paragraph{The three kinds of rotation.}
We classify rotations $\rho$ with $V_\rho\ne\varnothing$ according to  $|V_\rho|$ and $s(\rho)$.
Recall that $V_\rho$ consists of the labels whose first rotation is $\rho$.

\begin{itemize}
  \item An \emph{initial swap} is a swap $A=(u,v)$ that is the first rotation of both participants.
  Thus, $V_A=\{u,v\}$.
  \item An \emph{exceptional swap} is a rotation $B$ with $V_B=\{x\}$ and $s(B)=3$.
  By \Cref{lem:three-labels}, we may write $B=(u,x)$ after an initial swap $A=(u,v)$.
  We call $x$ \emph{exceptional} and $u$ its \emph{parent}.
  We say that $x$ is associated with $u$ and with the initial swap $A$.
  \item An \emph{ordinary rotation} is a rotation $\rho$ with $V_\rho\ne\varnothing$ that is neither an initial swap nor an exceptional swap.
  Its first group $V_\rho$ and the people with labels in $V_\rho$ are also called \emph{ordinary}.
\end{itemize}

Each parent has at most one associated exceptional person, because the exceptional swap must be the parent's second rotation. For an exceptional swap $B=(u,x)$ following $A=(u,v)$, we call $\{u,v,x\}$ its \emph{association triple}.

\begin{lemma}
\label{lem:reservation-supports}
  The following holds for $v$:
  \begin{enumerate}
	\item[\rm(i)] If $v$ is exceptional and $d_v\ge2$, then $s(\rho_{v,2})\ge4$.
	\item[\rm(ii)] If $v$ participates in an initial swap, has no associated exceptional person, and $d_v\ge2$, then $s(\rho_{v,2})\ge4$.	
	\item[\rm(iii)] If $\rho$ is ordinary  with $|V_\rho|\in\{1,2\}$, then $s(\rho)\ge4$.
\end{enumerate}
\end{lemma}
\begin{proof}
  We describe the rotations using men.
  The same arguments apply for women.
	
  For (i), let $x$ be exceptional.
  Write $A=(u,w)$ for its associated initial swap and $B=(u,x)$ for its first rotation.
  By \Cref{lem:three-labels}, $\past B=\{A,B\}$.
  After these two swaps, $u$ has had all three women with labels in $\{u,w,x\}$, so he cannot participate again within these labels.
  The only remaining possible rotation is $(w,x)$, but this would return $w$ to his initial partner.
  Thus, if $d_x\ge2$, the rotation $\rho_{x,2}$ must have support containing at least one further label.
  Hence, $s(\rho_{x,2})\ge4$.
	
  For (ii), suppose $v$ participates in an initial swap and $d_v\ge2$.
  By \Cref{lem:support}, $s(\rho_{v,2})\ge3$.
  If equality holds, eliminate $\past\rho_{v,2}$ first.
  Write the initial swap as $A=(u,v)$ and the support as $\{u,v,x\}$.
  If $x$ participates before $\rho_{v,2}$, its first rotation must be $(u,x)$, since $v$'s next rotation is $\rho_{v,2}$.
  The argument for (i) then rules out any further rotation using only these three labels, a contradiction.
  Thus $x$ first participates at $\rho_{v,2}$, and $V_{\rho_{v,2}}=\{x\}$.
  By \Cref{lem:three-labels}, $\rho_{v,2}$ is an exceptional swap with parent $v$.
  This contradicts the assumption that $v$ has no associated exceptional person.
  Therefore, $s(\rho_{v,2})\ge4$.
	
  For (iii), let $\rho$ be ordinary  with $|V_\rho|\in\{1,2\}$.
  If $s(\rho)=2$, no earlier rotation is possible, since it would swap the same two people and $\rho$ would restore their initial partners.
  Hence, $\rho$ is an initial swap, contrary to $\rho$ being ordinary.
  Suppose instead that $s(\rho)=3$.
  If $|V_\rho|=1$, then $\rho$ is an exceptional swap.
  If $|V_\rho|=2$, the third label must have participated in an earlier rotation, since it is not in $V_\rho$.
  But that rotation cannot involve either label in $V_\rho$, leaving only one possible participant.
  This is impossible.
  Thus, $s(\rho)\ge4$.
\end{proof}

For each active person $v$, define $\beta_v$ as $H_{d_v}$ plus the portions of their first two occurrence terms in $\Omega$ specified in \Cref{table:amountB_v}.
In the table, $d=d_v$ and $k$ is the size of $v$'s first group.
Here, $\ind_P$ equals $1$ if the condition $P$ holds, and $0$ otherwise.

\begin{table}[htbp]
	\centering
	\small
	\begin{tabular}{@{}p{0.35\linewidth}p{0.2\linewidth}@{}}
		\toprule
		Person & Reserved amount $\beta_v$\\
		\midrule
		Ordinary, $k\ge3$
		& $H_d+h_1$\\
		Ordinary, $k=1,2$
		& $H_d+2h_1+h_2\ind_{d\ge2}$\\
		Exceptional
		& $H_d+h_1+h_2\ind_{d\ge2}$\\
		Participant in an initial swap with an associated
		exceptional person
		& $H_d$\\
		Participant in an initial swap without an associated
		exceptional person
		& $H_d+h_2\ind_{d\ge2}$\\
		\bottomrule
	\end{tabular}
	\caption{The amounts reserved for each person.
		The terms added to $H_d$ are taken from that person's
		first two occurrence terms in $\Omega$.}
	\label{table:amountB_v}
\end{table}

The terms involving $h_1$ are taken from the first occurrence term $(s(b_v)-2)h_1$.
An exceptional person has $s(b_v)=3$.
An ordinary person has $s(b_v)\ge3$ when $k\ge3$, and $s(b_v)\ge4$ when $k=1,2$.
Thus these first occurrence terms supply the amounts specified in the table.
Whenever the table also reserves $h_2$, the person has a second rotation whose support has size at least four, by \Cref{lem:reservation-supports}.
Their second occurrence term $(s(\rho_{v,2})-3)h_2$ therefore supplies this amount.

Each reservation uses only that person's own terms.
Consequently, the total amount left unreserved is
\begin{equation}
\label{eq:unused-surplus}
	U=\sum_v H_{d_v}+\Omega-\sum_v\beta_v\ge0 \enspace .
\end{equation}
This includes all third and later occurrence terms, together with the unused parts of the first two.
For each ordinary rotation $\rho$, we will prove that $\sum_{v\in V_\rho}\beta_v$ covers its cost $|V_\rho|\lambda(p(\rho))$.
For each initial swap, we will consider its two participants together with their associated exceptional people.
We will bound how much their total cost can exceed the sum of their reserved amounts.
Later sections will combine the two sexes and cover any remaining excess using $U$ and amounts left over from ordinary groups.
This will establish \eqref{eq:summed-criterion}.

We now show that the reserved amounts cover the costs of ordinary groups.
We first extend~\eqref{cond:degree} to degree one.


\begin{lemma}
\label{lem:basic-degree-bounds}
  Every admissible sequence satisfies
  \begin{equation}
  \label{eq:scalar-all-degrees}
    H_d+h_2\ind_{d\ge2}\ge\lambda(2h_d)\qquad(d\ge1) \enspace .
  \end{equation}
\end{lemma}
\begin{proof}
  For $d\ge2$, this is \eqref{cond:degree}.
  For $d=1$, \eqref{cond:first} and $\lambda(\ln2)=\ln2$ give $\lambda(2h_1)\le\ln2\le2h_1=H_1$.
\end{proof}

\begin{lemma}
\label{lem:ordinary-cost}
  For an ordinary rotation $\rho$ with $|V_\rho|=k$, it holds
  \begin{equation*}
    k\lambda(p(\rho))\le\sum_{v\in V_\rho}\beta_v \enspace .
  \end{equation*}
\end{lemma}
\begin{proof}
  By \eqref{eq:group-convexity}, it suffices to show that $\lambda(kh_{d_v})\le\beta_v$ for every $v\in V_\rho$.
  Fix $v\in V_\rho$ and write $d=d_v$.
  If $k\ge3$, then
  \begin{equation*}
    \lambda(kh_d) \le\lambda(2h_d)
                  \le H_d+h_2\ind_{d\ge2}\le H_d+h_1 \enspace .
  \end{equation*}
  If $k=1,2$, then
  \begin{equation*}
	\lambda(kh_d) \le \lambda(h_d)
                  \le \lambda(2h_d)+\ln2
                  \le H_d+h_2\ind_{d\ge2}+2h_1 \enspace .
  \end{equation*}
  In each case, the final expression is $\beta_v$ as defined in \Cref{table:amountB_v}.
\end{proof}

Define the unused amount of an ordinary group $V_\rho$ by
\begin{equation}
\label{eq:ordinary-saving}
  \Delta_\rho=\sum_{v\in V_\rho}\beta_v-|V_\rho|\lambda(p(\rho))\ge0 \enspace .
\end{equation}
This is the part of its reserved amount that is left after its actual cost is covered.

Let $\mathcal P$ be the set of initial swaps, and let $\mathcal O$ be the set of ordinary rotations.
For $A=(u,v)\in\mathcal P$, let $X_A$ be the set of exceptional labels associated with $A$.
By \Cref{lem:three-labels}, $|X_A|\le2$.
Put
\begin{equation*}
  K_A=\{u,v\}\cup X_A \enspace .
\end{equation*}
The first groups partition $V$.
For each initial swap $A$, the set $K_A$ is the union of its first group and all exceptional first groups associated with $A$.
Each exceptional first group is associated with exactly one initial swap.
The remaining first groups are those of ordinary rotations.
Thus
\begin{equation}
\label{eq:partition}
  \{K_A:A\in\mathcal P\}\cup\{V_\rho:\rho\in\mathcal O\}
\end{equation}
is a partition of $V$.
For each $A\in\mathcal P$, define
\begin{equation*}
  D_A^* = 2\lambda(p^*(A)) + \sum_{x\in X_A}\lambda(p^*(b_x)) - \frac12\sum_{v\in K_A}(\beta_v^m+\beta_v^w) \enspace .
\end{equation*}
Thus $D_A^*$ is the total cost of the labels in $K_A$ minus the average of the reserved sums for men and women.

\subsection{Proof of the branching criterion}
\label{sec:BC-proof}
The partition \eqref{eq:partition} reduces the branching
criterion to the estimate in \Cref{prop:covering-budget} below.
For $s\in\{m,w\}$, let
\begin{equation*}
  U_s=\sum_{v\in V}H_{d_v^s}+\Omega_s-\sum_{v\in V}\beta_v^s
\end{equation*}
be the unreserved occurrence surplus, as in
\eqref{eq:unused-surplus}.

\begin{proposition}
\label{prop:covering-budget}
  For every admissible sequence and every $Q\in\mathcal Q$,
  \begin{equation*}
    \sum_{A\in\mathcal P}D_A^* \le\frac{U_m+U_w}{2} +\frac12\sum_{\rho\in\mathcal O}
                 (\Delta_\rho^m+\Delta_\rho^w) \enspace .
  \end{equation*}
\end{proposition}

\begin{proof}[\bfseries Proof of \Cref{prop:summed-criterion} assuming
\Cref{prop:covering-budget}]
  For an ordinary first group, convexity and the definition of its saving give
  \begin{equation*}
    |V_\rho|\lambda(p^*(\rho)) \le\frac12\sum_{v\in V_\rho}(\beta_v^m+\beta_v^w)
       -\frac12(\Delta_\rho^m+\Delta_\rho^w) \enspace .
  \end{equation*}
  For an initial-swap block, its cost is its averaged reserved amount plus $D_A^*$.
  Summing over the partition \eqref{eq:partition}, and using \Cref{lem:upset-sum}, therefore yields
  \begin{equation*}
    \sum_{v\in V}\bigl[\lambda(p^*(b_v))-q^*(b_v)\bigr] \le \sum_{A\in\mathcal P}D_A^*
       -\frac{U_m+U_w}{2}-\frac12\sum_{\rho\in\mathcal O}                   (\Delta_\rho^m+\Delta_\rho^w) \le 0 \enspace .
  \end{equation*}
  The last inequality is \Cref{prop:covering-budget}.
  This is precisely \eqref{eq:summed-criterion}.
\end{proof}

The next three sections prove \Cref{prop:covering-budget}.
We first bound each block's deficit, then retain the slack of low-degree parents and describe the matching constraints on four labels.
The simultaneous allocation in \Cref{sec:credit} covers every remaining positive deficit without reusing contributions.

\section{Costs of initial and exceptional swaps}
\label{sec:costs}
We first bound how much the total cost of the labels in $K_A$
can exceed the average of their reserved amounts for men and women.
For $s\in\{m,w\}$, define
\begin{equation*}
  N_A^s=\bigl|\{x\in X_A:d_x^s\ge2\}\bigr| \enspace .
\end{equation*}
We prove that $D_A^*\le h_2(N_A^m+N_A^w)/2$.
In \Cref{sec:visited37}, we improve this bound when a parent has small degree.
Finally, in \Cref{sec:credit}, we show that the amounts left unused by ordinary groups and by the first three occurrences are enough to cover every positive $D_A^*$, without using any amount twice.
This proves \Cref{prop:covering-budget}.

\medskip

We first derive some consequences of admissibility that will be used in our arguments.
For an exceptional label $v$ with $d_v=c$, we will first bound $\lambda(p(b_v))-A_c$, with $A_c$ as in \Cref{def:admissible}.
Compared with the reserved amount $\beta_v$, $A_c$ includes an additional $h_1$ when $c=1$ and an additional~$h_2$ when $c\ge2$.
We will account for these additional amounts later.

For $d\ge1$ and $a\ge2$, put
\begin{equation*}
  \sigma_d=H_d-\lambda(2h_d),\quad\text{and}\quad g_a=h_{a-1}+h_a \enspace .
\end{equation*}
In fact, $g_a$ is the decrease in a degree-$a$ person's potential when their first two
rotations are deleted.
Define,
\begin{equation*}
  \Phi(t)=\sup_{c\ge1}\{\lambda(t+h_c)-A_c\}\qquad\text{for }t\ge0 \enspace .
\end{equation*}
This supremum is finite.
Indeed, for $c\ge2$,
\begin{equation*}
  \lambda(t+h_c)-A_c \le\lambda(h_c)-\lambda(2h_c)-h_1-h_2
                     \le\ln2-h_1-h_2
\end{equation*}
by \eqref{cond:degree}; the term with $c=1$ is also finite.
Deleting $\past b_v$ removes exactly one of their rotations, so their contribution to $p(b_v)$ is $h_c$.
If the sum of the other people's contributions is at least $t$, then $p(b_v)\ge t+h_c$. Since $\lambda$ is decreasing,
\begin{equation}
\label{eq:phi-bound}
  \lambda(p(b_v))-A_c	\le\lambda(t+h_c)-A_c
                      \le\Phi(t)\enspace .
\end{equation}
This gives a bound that does not depend on $c$.
The function $\Phi$ is decreasing and convex, since each function $t\mapsto\lambda(t+h_c)-A_c$ has these properties.

Condition~\eqref{cond:exceptional} gives $\Phi(g_a+h_a)\le\sigma_a$ for $a\ge3$.
The following lemma also gives bounds for $a=2$, using only \eqref{cond:first} and \eqref{cond:degree}.

\begin{lemma}
\label{lem:costs}
  Every admissible sequence satisfies the following:
  \begin{enumerate}[label=\textup{(\roman*)},ref=\thelemma\,(\roman*)]
    \item\label[lemma]{lem:costs-ii} For every $a\ge2$ it holds $\Phi(g_a+h_a)\le\sigma_a-h_1\ind_{a=2}$.
    \item\label[lemma]{lem:costs-iii} $\Phi(g_2)\le\sigma_2-h_1$.
  \end{enumerate}
\end{lemma}
\begin{proof}
  Write $u=h_1$ and $b=h_2$.
  By \Cref{lem:basic-degree-bounds}, $\sigma_d\ge-b$ for $d\ge2$ and $\sigma_1\ge0$.
  For $c\ge2$, monotonicity and \eqref{cond:degree} give
  \begin{equation*}
    \lambda(u+b+h_c)-A_c \le\lambda(2h_c)-A_c
                         \le-u-b \enspace .
 \end{equation*}
  For $c=1$, \eqref{cond:first} gives
  \begin{equation*}
    \lambda(2u+b)-4u\le\lambda(2u)-4u\le-2u\le-u-b \enspace .
  \end{equation*}
  Taking the supremum over $c$ gives $\Phi(g_2)\le-u-b\le\sigma_2-u$, proving (ii).
  The case $a=2$ of (i) follows because $g_2+h_2\ge g_2$ and $\Phi$ is decreasing.
  For $a\ge3$, rearrange \eqref{cond:exceptional} and take the supremum over $c$.
\end{proof}

We will use the following consequence of convexity.
Since both $(g_d)$ and $(h_d)$ are decreasing, every convex function $f$ satisfies
\begin{equation}
\label{eq:convex-comparison}
  f(g_i+h_j)+f(g_j+h_i)
  \le f(g_i+h_i)+f(g_j+h_j)
  \qquad\text{for }i,j\ge2.
\end{equation}
Indeed, the two arguments on the left lie between the two arguments on the right and have the same sum.
Applying convexity to each term and adding proves the inequality.

\begin{lemma}
\label{lem:balances}
  For every initial swap $A$, it holds
  \begin{equation*}
	D_A^*\le\frac{h_2}{2}\bigl(N_A^m+N_A^w\bigr) \enspace .
  \end{equation*}
\end{lemma}
\begin{proof}
  Fix $A=(u,v)$.
  For $x\in X_A$, let $a_x^m$ and $a_x^w$ be the degrees of its parent in the men's and women's descriptions, respectively.
  We first use the men's occurrence chains and omit the index $m$.
  Write $a=d_u$ and $b=d_v$.
  We prove
  \begin{equation}
    \label{eq:1lambda(p_A)}
	2\lambda(p(A))+\sum_{x\in X_A}\lambda(p(b_x))
		\le\beta_u+\beta_v+\sum_{x\in X_A}A_{d_x}
		-h_1\sum_{x\in X_A}\ind_{a_x=2} \enspace .
  \end{equation}
  Since $\past A=\{A\}$, we have $p(A)=h_a+h_b$.
  Convexity gives
  \begin{equation*}
	2\lambda(p(A)) \le\lambda(2h_a)+\lambda(2h_b)
	= H_a+H_b-\sigma_a-\sigma_b \enspace .
  \end{equation*}
  If $x$ is associated with $u$ and $c=d_x$, then $p(b_x)\ge g_a+h_b+h_c$, and therefore
  \begin{equation*}
	\lambda(p(b_x))-A_c \le\lambda(g_a+h_b+h_c)-A_c \le\Phi(g_a+h_b) \enspace .
  \end{equation*}
  The corresponding bound for an association with $v$ has $a,b$ interchanged.
  We also use
  \begin{equation*}
	\sigma_d+h_2\ind_{d\ge2}\ge0\qquad\text{for }d\ge1,
  \end{equation*}
  which follows from \eqref{eq:scalar-all-degrees}.
	
  If $X_A=\varnothing$, then \eqref{eq:scalar-all-degrees} gives
  \begin{equation*}
	2\lambda(p(A)) \le\lambda(2h_a)+\lambda(2h_b) \le H_a+H_b+h_2(\ind_{a\ge2}+\ind_{b\ge2})
	=\beta_u+\beta_v \enspace .
  \end{equation*}
  This proves \eqref{eq:1lambda(p_A)} in this case.
	
  Suppose $X_A=\{x\}$.
  We may assume that $x$ is associated with $u$.
  Then
  \begin{equation*}
	\beta_u=H_a,\quad\text{and}\quad \beta_v=H_b+h_2\ind_{b\ge2} \enspace .
  \end{equation*}
  If $a\ge3$, put $\ell=\max(a,b)$ and $j=\min(a,b)$.
  Monotonicity of the weights and of $\Phi$, followed by \Cref{lem:costs-ii}, gives
  \begin{equation*}
    \Phi(g_a+h_b)\le\Phi(g_\ell+h_\ell)\le\sigma_\ell\enspace .
  \end{equation*}
  Since $j=1$ when $b=1$, and $j\ge2$ otherwise, we have $\sigma_j+h_2\ind_{b\ge2}\ge0$.
  Hence
  \begin{equation*}
	\Phi(g_a+h_b)
	\le\sigma_a+\sigma_b+h_2\ind_{b\ge2} \enspace .
  \end{equation*}
  If $a=2$, \Cref{lem:costs-iii} gives
  \begin{equation*}
    \Phi(g_2+h_b)\le\Phi(g_2)\le\sigma_2-h_1
	\le\sigma_2+\sigma_b+h_2\ind_{b\ge2}-h_1 \enspace .
  \end{equation*}
  Thus, in both cases,
  \begin{equation*}
	2\lambda(p(A))+\lambda(p(b_x)) \le\beta_u+\beta_v+A_{d_x}-h_1\ind_{a=2} \enspace .
  \end{equation*}
  This proves \eqref{eq:1lambda(p_A)} when $X_A=\{x\}$, since $a_x=a$.
	
  Finally, suppose $|X_A|=2$, with one exceptional label associated with each parent. Applying~\eqref{eq:convex-comparison} to $\Phi$ and then using \Cref{lem:costs-ii} gives
  \begin{equation*}
    \Phi(g_a+h_b)+\Phi(g_b+h_a)
    \le\Phi(g_a+h_a)+\Phi(g_b+h_b)\le\sigma_a+\sigma_b
       -h_1(\ind_{a=2}+\ind_{b=2}).
  \end{equation*}
  Combining this with the bound for $2\lambda(p(A))$ proves \eqref{eq:1lambda(p_A)}, since $\beta_u=H_a$ and $\beta_v=H_b$.

  The same argument applies to the women's occurrence chains, using their degrees and parents. Thus we have both the men's and women's versions of \eqref{eq:1lambda(p_A)}.
  For $s\in\{m,w\}$ and $x\in X_A$, the definitions give
  \begin{equation*}
	A_{d_x^s} =\beta_x^s+h_2\ind_{d_x^s\ge2}+h_1\ind_{d_x^s=1} \enspace .
  \end{equation*}
  Substituting this identity and $N_A^s=\sum_{x\in X_A}\ind_{d_x^s\ge2}$ into the corresponding version of \eqref{eq:1lambda(p_A)} gives
  \begin{equation}
  \label{eq:2lambda(p_sA)}
		2\lambda(p_s(A))+\sum_{x\in X_A}\lambda(p_s(b_x))
		\le\sum_{v\in K_A}\beta_v^s+h_2N_A^s
		+h_1\sum_{x\in X_A}
		(\ind_{d_x^s=1}-\ind_{a_x^s=2}) \enspace .
  \end{equation}	
  Fix $x\in X_A$.
  Interchanging $u,v$ if necessary, write $b_x=B=(u,x)$ after $A=(u,v)$.
  After $B$, man~$x$ is matched with woman $v$, and man $u$ is matched with woman~$x$.
  This is the first occurrence of man~$x$ and woman $x$, and the second occurrence of man $u$ and woman $v$.	
  Each pair remains matched until the next rotation involving either partner.
  Consequently,
  \begin{equation*}
	d_x^m=1\quad\Longleftrightarrow\quad d_v^w=2,
	\qquad
	d_x^w=1\quad\Longleftrightarrow\quad d_u^m=2 \enspace .
  \end{equation*}
  Since $a_x^m=d_u^m$ and $a_x^w=d_v^w$, these equivalences give
  \begin{equation*}
	(\ind_{d_x^m=1}-\ind_{a_x^m=2})
	+(\ind_{d_x^w=1}-\ind_{a_x^w=2})=0 \enspace .
  \end{equation*}
  Thus, the $h_1$ terms in \eqref{eq:2lambda(p_sA)} cancel when the inequalities for men and women are added.
  By convexity,
  \begin{equation*}
    \lambda(p^*(\rho)) \le\frac{\lambda(p_m(\rho))+\lambda(p_w(\rho))}{2} \enspace .
  \end{equation*}
  Apply this to $A$ and each $b_x$, and average \eqref{eq:2lambda(p_sA)} for $s=m$ and $s=w$.
  Since the $h_1$ terms cancel, we obtain
  \begin{equation*}
    D_A^* \le\frac12\sum_{s\in\{m,w\}}\left(2\lambda(p_s(A))
		+\sum_{x\in X_A}\lambda(p_s(b_x))\right)
		-\frac12\sum_{v\in K_A}(\beta_v^m+\beta_v^w)
		\le\frac{h_2}{2}\bigl(N_A^m+N_A^w\bigr) \enspace .\qedhere
  \end{equation*}
\end{proof}

\section{Small supports and low-degree parents}
\label{sec:visited37}
We prove two facts needed in \Cref{sec:credit}.
For an initial swap $A=(u,v)$ with two associated exceptional labels, we show that $D_A^*\le0$ if any of $d_u^m,d_v^m,d_u^w,d_v^w$ is at most three.
We also examine partner changes among four labels.
Since nobody can return to a previous partner, some pairs cannot change again without involving an additional label.
We identify when this happens and show that the support must then grow.
These facts help us cover the remaining positive values of $D_A^*$.

The proofs use \eqref{cond:transfer} and \eqref{cond:stop}.
Throughout this section, write $a=h_1$, $b=h_2$, and $c=h_3$.

\subsection{Retaining the slack of a parent}
Put $t_d=g_d+h_d=h_{d-1}+2h_d$ for $d\ge2$, and define
\begin{equation*}
  \Phi_2(t)=\sup_{j\ge2}\{\lambda(t+h_j)-A_j\} \enspace .
\end{equation*}
Like $\Phi$, this is decreasing and convex, and $\Phi_2\le\Phi$.
It is the appropriate bound when the exceptional people in one description all have degree at least two.
For $d\ge2$, define
\begin{equation*}
  \eta_d=\sigma_d-\Phi(t_d)-a\ind_{d=2},
\quad\text{and}\quad
\widehat\eta_d=\sigma_d-\Phi_2(t_d)-a\ind_{d=2} \enspace.
\end{equation*}
These are the amounts left after the estimates in
\Cref{lem:balances}.
Both are nonnegative by \Cref{lem:costs-ii}, and $\widehat\eta_d\ge\eta_d$ because $\Phi_2\le\Phi$.
By \Cref{lem:costs-ii}, both quantities are nonnegative.
They measure the part of a parent's degree contribution left over after the estimates in \Cref{lem:balances}.
\begin{lemma}
\label{lem:parent-slack}
  Every admissible sequence satisfies $\eta_2\ge b$, $\widehat\eta_2\ge3b$, and $\widehat\eta_3\ge4b$.
\end{lemma}
\begin{proof}
  We first note two elementary comparisons:
  \begin{equation}
  \label{eq:two-log-gaps}
    \lambda(2b)-\lambda(5b)>3b, \quad\text{and}\quad \lambda(2b)-\lambda(2b+2c)>2b \enspace .
  \end{equation}
  For $t>0$, we use the elementary bounds
  \begin{equation*}
    te^{-t/2}\le1-e^{-t}  \quad\text{and}\quad  1-e^{-t}\le\frac{t}{1+t/2} \enspace .
  \end{equation*}
  Taking negative logarithms gives
  \begin{equation*}
    \lambda(t)\le-\ln t+\frac t2 \quad\text{and}\quad \lambda(2b)\ge-\ln(2b)+b-\frac{b^2}{2},
  \end{equation*}
  where the second inequality also uses $\ln(1+b)\ge b-b^2/2$.
  Moreover, \eqref{cond:stop} gives $b\le4/25<1/6$ and $c/b\ge5/8$.
  The inequality $\ln x\ge2(x-1)/(x+1)$ for $x\ge1$ therefore yields
  \begin{align*}
     \lambda(2b)-\lambda(5b)-3b &\ge\frac67-\frac14-\frac1{72}-\frac12
                                  =\frac{47}{504}>0,\\
 \lambda(2b)-\lambda(2b+2c)-2b
 &\ge\frac{10}{21}-\frac1{10}-\frac1{72}-\frac13
   =\frac{73}{2520}>0 \enspace .
\end{align*}

  For $T\ge b$, the function $z\mapsto\lambda(T+z)-\lambda(2z)$ is increasing on $0<z\le b$: its derivative is at least $-\lambda'(2z)>0$.
  Condition \eqref{cond:degree} consequently gives
  \begin{equation}\label{eq:phi2-simple}
    \Phi_2(T)\le\lambda(T+b)-\lambda(2b)-a-b \enspace .
  \end{equation}
  Since $t_2=a+2b\ge4b$ and $\sigma_2\ge-b$, the first comparison in \eqref{eq:two-log-gaps} proves $\Phi_2(t_2)\le\sigma_2-a-3b$.
  Similarly, $t_3=b+2c$ and $\sigma_3\ge-b$ give $\Phi_2(t_3)\le-a-3b\le\sigma_3-4b$, using $a\ge2b$.
  Finally, the term $j=1$ omitted from $\Phi_2(t_2)$ is
  \begin{equation*}
    \lambda(2a+2b)-4a\le-2a \le\sigma_2-a-b,
  \end{equation*}
  because $\lambda(2a)\le2a$, $\sigma_2\ge-b$, and $a\ge2b$.
  Taking the maximum with the bound for $\Phi_2(t_2)$ proves the remaining assertion.
\end{proof}

\begin{lemma}
\label{lem:low-parent}
  Suppose $|X_A|=2$.
  If $D_A^*>0$, then
  \begin{equation*}
    d_u^m,d_v^m,d_u^w,d_v^w\ge4 \enspace .
  \end{equation*}
\end{lemma}
\begin{proof}
  All four parent degrees are at least two.
  In one sex, with parent degrees $i,j$, use $\Psi=\Phi_2$ if both exceptional degrees are at least two, and otherwise use $\Psi=\Phi$. Applying \eqref{eq:convex-comparison} to $\Psi$ gives
  \begin{equation*}
    \Psi(g_i+h_j)+\Psi(g_j+h_i) \le\Psi(t_i)+\Psi(t_j) \enspace .
  \end{equation*}
  As in the two-association estimate of \Cref{lem:balances}, this yields
  \begin{equation*}
   2\lambda(p(A))+\sum_{x\in X_A}\lambda(p(b_x)) \le H_i+H_j+\sum_{x\in X_A}A_{d_x}
      -\bigl(\sigma_i-\Psi(t_i)\bigr)
      -\bigl(\sigma_j-\Psi(t_j)\bigr) \enspace .
  \end{equation*}
  When both exceptional degrees are at least two, we may use either $\Phi$ or $\Phi_2$.
  We now substitute
  \begin{equation*}
    \sigma_d-\Phi(t_d)=a\ind_{d=2}+\eta_d
    \quad\text{and}\quad
  \sigma_d-\Phi_2(t_d)=a\ind_{d=2}+\widehat\eta_d \enspace .
  \end{equation*}
  For each exceptional swap, its degree-one indicators in the two sexes equal the opposite parent degree-two indicators, as proved in \Cref{lem:balances}.
  Thus the $a$ terms cancel when the two inequalities are added.
  If $z$ of the four parent degrees equal two, exactly $4-z$ exceptional people, counting both sexes, have degree at least two.
  For each sex $s$, let $L_s$ be the total amount retained for the two parents: set $L_s=\eta_{d_u^s}+\eta_{d_v^s}$ when using $\Phi$, and $L_s=\widehat\eta_{d_u^s}+\widehat\eta_{d_v^s}$ when using $\Phi_2$.
  By convexity of $\lambda$, averaging the estimates for men and women gives
  \begin{equation}
  \label{eq:retained-parent-slack}
    D_A^*\le\frac b2(4-z)-\frac{L_m+L_w}{2} \enspace .
  \end{equation}
  If $z\ge2$, use $\Phi$ for both men and women.
  \Cref{lem:parent-slack} makes the right-hand side at most \mbox{$b(4-2z)/2\le0$}.
  If $z=1$, both exceptional degrees in the sex of that degree-two parent are at least two: the opposite sex has no degree-two parent.
  Use $\Phi_2$ in that sex to obtain a saving of at least $3b$, again making \eqref{eq:retained-parent-slack} nonpositive.
  Finally, if $z=0$, use $\Phi_2$ in both sexes.
  Any parent degree three then supplies at least $4b$, which covers the entire first term. Positivity therefore forces all four degrees to be at least four.
\end{proof}

\subsection{When four labels cannot support a further change}
Suppose a person's last rotation in $D$ has support $K$ and their
current pair is fixed within $K$. If their next rotation $\tau$ exists,
then $K\subseteq S(\tau)$ by monotonicity of supports.
If $S(\tau)=K$, we can eliminate $D$ first and then the remaining
rotations of the downset $D\cup\past\tau$. This separates the pair
at $\tau$ using only labels in $K$, a contradiction.
Hence $S(\tau)$ strictly contains $K$.

Call a current pair \emph{fixed within $K$} if it cannot separate through such changes.
This happens if one partner has already been matched to every possible partner in $K$.
It also happens if all partners that person has not yet had belong to pairs already known to be fixed within $K$.
We will use this reasoning repeatedly.
If a person's last rotation in $D$ has support $K$ and their current pair is fixed within $K$, their next rotation must have support strictly containing $K$.

The labels involved in a downset form a union of whole first groups.
Also, association triples belonging to different initial swaps are disjoint, since they lie in
the disjoint sets $K_A$.
We use these facts to determine the possible fourth label in the following lemma.
\begin{lemma}
\label{lem:four-matching}
  Let $A=(u,v)$ and $B=(u,x)$ be initial and exceptional swaps.
  Suppose the second rotation $C$ of $x$ has support $K=\{u,v,x,y\}$, and put $\gamma=b_y$.
  \begin{enumerate}[label=\textup{(\roman*)}]
    \item\label{item:four-at-once} If $\gamma=C$, man $y$'s pair after $C$ is fixed within $K$.
      Thus, $y$ stops or its second support has size at least five.
    \item\label{item:four-before} If $y$ is ordinary and $\gamma\ne C$, the matching after  $\past C$, in the order $u,v,x,y$, has partner labels  $(y,x,u,v)$. All its pairs are fixed within $K$, and each man's  last occurrence in $\past C$ has support~$K$.
   \item\label{item:four-exceptional} If $y$ is exceptional, it belongs to $A$, $\gamma=E=(v,y)$, and $C=(x,y)$.
     The pairs of $x,y$ after $C$ are fixed within $K$.
     Moreover, if an ordinary label $z\notin K$ has first support of size four contained in $K\cup\{z\}$, then $z$ stops or its second support has size at least five.
\end{enumerate}
\end{lemma}
\begin{proof}
  Eliminate $A,B$ first.
  The partner labels of $u,v,x,y$ are $(x,u,v,y)$, and no further rotation on $\{u,v,x\}$ is possible.
  Woman $v$, held by $x$, has visited men $v,u,x$.

  If $\gamma=C$, she must move to $y$, exhausting her partners in $K$ and fixing their pair. 
  This proves~(i).

  If $y$ is ordinary and moves before $C$, its first rotation omits $x$ and must contain $u$; otherwise it would be the exceptional swap $(v,y)$.
  Man $u$ can receive only woman $y$, so his pair is now fixed.
  At~$C$, man $x$ must receive woman $u$, and woman $v$ must go to $y$.
  The remaining partner of $v$ is woman~$x$.
  Man $u$ and woman $v$ have exhausted all partners in $K$.
  Fixing their pairs also fixes $x$'s pair and hence the remaining pair.
  The first rotation of $y$ follows $B$ on $u$'s chain, so has support $K$.
  After it, any rotation before $C$ must contain $y$ (it cannot use the fixed $u$ or the not-yet-moving $x$).
  Finally, $v$ must move at or after that first rotation, because his partner changes from $u$ to $x$.
  All four last occurrences therefore have support $K$.

  If $y$ is exceptional, the union-of-first-groups observation forces its initial swap to be $A$, with parent $v$.
  After $A,B,E$, the women held by $x,y$ are $v,u$; each can move within $K$ only to the other man.
  Thus $C=(x,y)$, after which both women have visited every man in $K$, fixing these pairs. For the final assertion, the first support of $z$ consists of $z,u,v$ and one of $x,y$,
say $x$.
  The next rotations of $v,x$ require the missing label $y$, so neither can move in this support.
  The first rotation of $z$ is consequently $(u,z)$ after $A,B$.
  It exhausts $u$'s partners, fixing their pair.
  Man $z$ then holds woman~$x$, whose only unvisited man in this support is the fixed $v$.
  Thus $z$'s pair is fixed as well.
\end{proof}

We also record how this argument detects a first support of size five.
If all pairs are fixed after a downset supported on four labels $K$, and every man's last occurrence there has support $K$, then any continuation with just one further label $z$ starts with $b_z$ of support $K\cup\{z\}$.
Indeed, the first further rotation must contain $z$ and an old participant whose preceding occurrence already has support $K$.
This first group is an ordinary singleton.

\section{Covering the remaining costs}
\label{sec:credit}
We now prove \Cref{prop:covering-budget} by showing that the amounts left unused are enough to cover every positive $D_A^*$.
We first use contributions from first occurrences in ordinary groups whose supports have size at least five.
For the remaining cases, we use amounts left over from ordinary groups and from second and third occurrences.
We show that each amount is used at most once.
Write $b=h_2$ and $c=h_3$.

\begin{lemma}
\label{lem:large-first-cover}
  Let $\rho$ be an ordinary rotation with $s(\rho)\ge5$.
  Let $\mathcal A_\rho$ be the set of initial swaps having an association triple contained in $S(\rho)$.
  Then
  \begin{equation*}
    \sum_{A\in\mathcal A_\rho}\max\{D_A^*,0\} \le |V_\rho|\bigl(s(\rho)-4\bigr)h_1 \enspace .
  \end{equation*}
  For each sex, the right-hand side is available from the unreserved first-occurrence contributions of~$V_\rho$.
\end{lemma}
\begin{proof}
  Put $k=|V_\rho|$, $s=s(\rho)$, and $m=|\mathcal A_\rho|$.
  Choose one association triple contained in $S(\rho)$ for each $A\in\mathcal A_\rho$.
  These triples are disjoint and avoid $V_\rho$, so $3m+k\le s$.
  We claim that $k(s-4)\ge m$. For $m\le1$, this follows from $k\ge1$ and $s\ge5$.
  For $m\ge2$,
  \begin{equation*}
    s-4\ge3m+k-4\ge3m-3\ge m,
  \end{equation*}
  which also proves the claim.
  By \Cref{lem:balances}, every initial swap satisfies $D_A^*\le2h_2$.
  Using $h_1\ge2h_2$ from \eqref{cond:transfer}, we obtain
  \begin{equation*}
    \sum_{A\in\mathcal A_\rho}\max\{D_A^*,0\}
    \le2h_2m
    \le h_1m
    \le k(s-4)h_1 \enspace .
  \end{equation*}

  For either sex, each person in $V_\rho$ contributes $(s-2)h_1$ to $\Omega$ at their first occurrence.
  The reservation table uses at most $2h_1$ from this contribution.
  Thus the first occurrences of $V_\rho$ leave at least $k(s-4)h_1$ unreserved, as required.
\end{proof}

Let $\mathcal P_5$ be the set of initial swaps having an association triple contained in $S(\rho)$ for some ordinary rotation $\rho$ with $s(\rho)\ge5$.
Put $\mathcal P_0=\mathcal P\setminus\mathcal P_5$, and define
\begin{equation*}
  C_5=\sum_{A\in\mathcal P_5}\max\{D_A^*,0\} \enspace .
\end{equation*}
For each $A\in\mathcal P_5$, choose one such rotation $\rho$.
By \Cref{lem:large-first-cover}, the unused first-occurrence contributions of each chosen group are enough to cover all amounts assigned to it, separately for men and women.
Thus $C_5$ is covered without using any contribution twice.

We now consider the initial swaps in $\mathcal P_0$.
For each $A\in\mathcal P_0$, let $Z_A$ be the set of labels $z$ whose first rotation $b_z$ is ordinary and whose support $S(b_z)$ contains an association triple of $A$.
Since $A\notin\mathcal P_5$, each such support has size at most four.
The association triple uses three labels outside~$V_{b_z}$, so the support has size exactly four and $V_{b_z}=\{z\}$.
The sets $Z_A$ are disjoint for different $A$: a support of size four cannot contain two disjoint association triples.

For each $A\in\mathcal P_0$, consider one sex and define
\begin{equation*}
  E_x=\sum_{r=2}^{\min\{d_x,3\}}
       \bigl(s(\rho_{x,r})-4\bigr)h_r
  \qquad\text{for }x\in X_A.
\end{equation*}
Also set $T_z=\Delta_{b_z}$ for $d_z=1$, and when $d_z\ge2$:
\begin{equation*}
  T_z=\Delta_{b_z}
      +b\bigl(s(\rho_{z,2})-4\bigr)
  \qquad\text{for }z\in Z_A.
\end{equation*}
Put
\begin{equation*}
  R_A=\sum_{x\in X_A}E_x+\sum_{z\in Z_A}T_z.
\end{equation*}

\begin{lemma}
\label{lem:available-amounts}
  All quantities $E_x$ and $T_z$ are nonnegative.
  For either sex $s$,
  \begin{equation}
  \label{eq:disjoint-new-resources}
    C_5+\sum_{A\in\mathcal P_0}R_A^s \le U_s+\sum_{\rho\in\mathcal O}\Delta_\rho^s.
  \end{equation}
  Moreover, for each $z\in Z_A$:
  \begin{enumerate}[label=\textup{(\roman*)}]
    \item $T_z\ge b$ if $d_z=1$, or if $d_z\ge2$ and $s(\rho_{z,2})\ge5$.
    \item $T_z>2b$ if the proper past of $b_z$ contains a degree-two person of the sex under consideration.
  \end{enumerate}
\end{lemma}
\begin{proof}
An exceptional person's second support has size at least four, and their third support has size at least four by \Cref{lem:support}.
Thus, $E_x\ge0$.
For $z\in Z_A$, the first support has size four, so any second support also has size at least four.
Since $\Delta_{b_z}\ge0$, this gives $T_z\ge0$.
We next check that these amounts are available.
For an exceptional person $x$, the unreserved part of the second occurrence is $(s(\rho_{x,2})-4)h_2$.
The third occurrence contributes $(s(\rho_{x,3})-4)h_3$, none of which was reserved.
These are exactly the terms in $E_x$.
Similarly, $T_z$ uses the ordinary group's saving $\Delta_{b_z}$ and the unreserved part of $z$'s second occurrence.
The sets $X_A$ and $Z_A$ are disjoint for different $A$.
Also, each $z\in Z_A$ forms an ordinary singleton group.
Thus no occurrence contribution or ordinary saving is used twice.
The amount $C_5$ uses only first occurrences of ordinary groups with support size at least five, so it uses none of the contributions just listed.
This proves \eqref{eq:disjoint-new-resources}.

To prove \textup{(i)}, suppose first that $d_z=1$.
Then $\beta_z=4h_1$ and $p(b_z)\ge h_1$, so
\begin{equation*}
  T_z=\Delta_{b_z}
  \ge4h_1-\lambda(h_1)\ge b
\end{equation*}
by \eqref{cond:transfer}.
If $s(\rho_{z,2})\ge5$, the second term in $T_z$ is already at least $b$.

For \textup{(ii)}, write $a=h_1$ and $d=d_z$.
Deleting $\past b_z$ removes a rotation of the degree-two person, contributing at least $b$, as well as $z$'s first rotation, contributing $h_d$.
Hence $p(b_z)\ge b+h_d$.
If $d\ge2$, then $h_d\le b$, so $\lambda(b+h_d)\le\lambda(2h_d)$.
Using \eqref{cond:degree}, we obtain
\begin{equation*}
  T_z\ge\Delta_{b_z}
  \ge H_d+2a+b-\lambda(b+h_d)
  \ge2a>2b.
\end{equation*}
If $d=1$, then $T_z\ge4a-\lambda(a+b)$.
To show that this exceeds $2b$, define
\begin{equation*}
  f(t)=e^{-(4a-2t)}+e^{-(a+t)}.
\end{equation*}
The function $f$ is strictly convex.
Since $2a\ge\ln2$,
\begin{equation*}
  f(0)\le\frac14+\frac1{\sqrt2}<1
  \quad\text{and}\quad
  f(a)=2e^{-2a}\le1 \enspace .
  \end{equation*}
  Also, $0<b\le a/2<a$, so $f(b)<1$.
  Rearranging this inequality gives $4a-\lambda(a+b)>2b$, completing the proof.
\end{proof}
\begin{lemma}
\label{lem:small-core-cover}
  If $A\in\mathcal P_0$ and $D_A^*>0$, then for either sex
  \begin{equation*}
    R_A\ge bN_A \enspace .
  \end{equation*}
\end{lemma}
\begin{proof}
  Fix one sex and suppress the corresponding superscript.
  By \Cref{lem:available-amounts}, all terms in $R_A$ are nonnegative.
  Since $N_A\le2$, any $z\in Z_A$ with $T_z\ge2b$ already gives $R_A\ge bN_A$.
  We may therefore assume that $T_z<2b$ for every $z\in Z_A$.
  We use $2c\ge b$ from \eqref{cond:transfer}.

  \emph{Case 1. Two exceptional people have a common second rotation of support four.}
  Write $A=(u,v)$, $B=(u,x)$, and $E=(v,y)$ for the initial swap and its two exceptional swaps. By \Cref{lem:four-matching}\,(iii), the common second rotation is $C=(x,y)$, with support   $K_A=\{u,v,x,y\}$.
  In particular, $N_A=2$.

  After $C$, each of $x$ and $y$ is matched to a parent of the opposite sex for whom $C$ is the third rotation.
  If either $x$ or $y$ had no further rotation, that person's partner would also have no further rotation and would have degree three.
  This contradicts \Cref{lem:low-parent}, since $D_A^*>0$.
  Thus, both $x$ and~$y$ have a third rotation.
  Their pairs after $C$ are fixed within $K_A$ by \Cref{lem:four-matching}\,(iii), so both third supports strictly contain $K_A$.

  If both third supports have size at least six, then $E_x+E_y\ge4c\ge2b$, as required. Otherwise, one of the third rotations, say $\tau$, has support $K_A\cup\{z\}$ for some   $z\notin K_A$.
  We show that $z\in Z_A$.

  Since $b_z\preceq\tau$, we have $S(b_z)\subseteq K_A\cup\{z\}$.
  The four labels in $K_A$ first occur at $A$, $B$, or $E$, none of which contains $z$.
  Hence $V_{b_z}=\{z\}$, so $b_z$ is not an initial swap.
  If $z$ were exceptional, its associated initial swap would have to be $A$:
  any other initial swap would require two labels in $S(b_z)\setminus K_A$, which contains only $z$.
  But each parent in $A$ already has an associated exceptional person.
  Thus,~$b_z$ is ordinary.
  By \Cref{lem:reservation-supports}\,(iii), $s(b_z)\ge4$.
  Hence $S(b_z)$ contains at least three labels of $K_A$, including at least one of $x$ and $y$. The first rotation of that exceptional label precedes~$b_z$, so its entire association triple lies in $S(b_z)$.
  If $s(b_z)=5$, this ordinary rotation would put $A$ in~$\mathcal P_5$, contrary to  $A\in\mathcal P_0$.
  Thus $s(b_z)=4$ and $z\in Z_A$.
  The final assertion of \Cref{lem:four-matching}\,(iii) now gives either $d_z=1$ or  $s(\rho_{z,2})\ge5$.
  In either case, \Cref{lem:available-amounts}\,(i) gives $T_z\ge b$.
  Since both third supports have size at least five, $E_x+E_y\ge2c\ge b$.
  Therefore, $R_A\ge E_x+E_y+T_z\ge2b=bN_A$.

  \emph{Case 2. There is no common second rotation of support four.}
  Consider any $x\in X_A$ counted by~$N_A$, and let $C=\rho_{x,2}$.
  By \Cref{lem:reservation-supports}\,(i), $s(C)\ge4$.
  If $s(C)\ge5$, its contribution to $E_x$ is at least~$b$.
  We may therefore assume that $s(C)=4$.

  Name the parents so that $A=(u,v)$ and $b_x=(u,x)$, and write $S(C)=\{u,v,x,y\}$.
  Put $\gamma=b_y$. Since $\gamma\preceq C$, we have $S(\gamma)\subseteq S(C)$.
  The labels $u,v,x$ first occur at $A$ or $b_x$, neither of which contains $y$.
  Thus, $V_\gamma=\{y\}$.
  If $y$ were exceptional, \Cref{lem:four-matching}\,(iii) would make $C$ a common second
  rotation of $x$ and $y$, contrary to the assumption of this case.
  Hence $\gamma$ is ordinary, and \Cref{lem:reservation-supports}\,(iii) gives $4\le s(\gamma)\le s(C)=4$.
  It follows that $S(\gamma)=S(C)$.
  In particular, $S(\gamma)$ contains the association triple $\{u,v,x\}$, so $y\in Z_A$.

  If $\gamma=C$, then \Cref{lem:four-matching}\,(i) gives either $d_y=1$ or $s(\rho_{y,2})\ge5$. Thus
  \Cref{lem:available-amounts}\,(i) gives $T_y\ge b$.

  Suppose instead that $\gamma\ne C$.
  By \Cref{lem:four-matching}\,(ii), after eliminating $\past C$ all four pairs are fixed within $S(C)$, and the last rotation of each of the four people of the chosen sex has support $S(C)$.
  The proof of that lemma also gives $b_x\prec\gamma$.
  If $d_x=2$, the proper past of $\gamma=b_y$ would therefore contain the rotation $b_x$ of the degree-two person $x$.
  By \Cref{lem:available-amounts}\,(ii), this would give $T_y>2b$, contrary to our assumption.

  Thus, $x$ has a third rotation $\tau$.
  Since its pair after $C$ is fixed within $S(C)$, we have $S(C)\subsetneq S(\tau)$.
  Suppose $s(\tau)=5$, and write $S(\tau)=S(C)\cup\{z\}$.
  Eliminate $\past C$ first within $\past\tau$.
  By the observation following \Cref{lem:four-matching}, the first subsequent rotation is the ordinary first rotation $b_z$, with support
  $S(C)\cup\{z\}$.
  This support has size five and contains the association triple $\{u,v,x\}$, which would put~$A$ in~$\mathcal P_5$.
  Hence $s(\tau)\ge6$, and its contribution to $E_x$ is at least $2c\ge b$.

  We have obtained at least $b$ for each $x$ counted by $N_A$, either from its own $E_x$ or from a term~$T_y$ with $S(b_y)=\{u,v,x,y\}$.
  These terms are distinct.
  Indeed, if two distinct exceptional labels $x,x'$ used the same $T_y$, then $S(b_y)$ would contain both association triples $\{u,v,x\}$ and $\{u,v,x'\}$ as well as the ordinary label $y$.
  This would require five labels in a support of size four.
  Summing over the~$N_A$ exceptional people gives $R_A\ge bN_A$.
\end{proof}

\begin{proof}[\bfseries Proof of \Cref{prop:covering-budget}]
  Let $A\in\mathcal P_0$.
  If $D_A^*\le0$, then $D_A^*\le(R_A^m+R_A^w)/2$, since $R_A^m,R_A^w\ge0$.
  If $D_A^*>0$, \Cref{lem:balances,lem:small-core-cover} give
  \begin{equation*}
    D_A^* \le\frac{h_2}{2}(N_A^m+N_A^w) 
          \le\frac{R_A^m+R_A^w}{2} \enspace .
  \end{equation*}
  Thus this bound holds for every $A\in\mathcal P_0$.
  By the definition of $C_5$ we have $\sum_{A\in\mathcal P_5}D_A^*\le C_5$.
  Since~$\mathcal P_0$ and~$\mathcal P_5$ partition $\mathcal P$, summing gives
  \begin{equation*}
    \sum_{A\in\mathcal P}D_A^* \le C_5+\frac12\sum_{A\in\mathcal P_0}                  (R_A^m+R_A^w) \enspace .
  \end{equation*}
  Finally, averaging \eqref{eq:disjoint-new-resources} for men and women gives
  \begin{equation*}
    C_5+\frac12\sum_{A\in\mathcal P_0}(R_A^m+R_A^w)\le\frac{U_m+U_w}{2}     +\frac12\sum_{\rho\in\mathcal O}(\Delta_\rho^m+\Delta_\rho^w) \enspace .
  \end{equation*}
  Combining these inequalities proves the proposition.
\end{proof}

\section{Finding the weights by finite optimization}
\label{sec:numbers}
We have proved that every admissible sequence makes $F_h$ a branching potential. It remains to construct one that also satisfies $S=\sum_{d\ge1}h_d<\ln2.81$.
In this section, we reduce this task to a finite convex optimization problem. We then give one feasible solution which serves as explicit weights and verify that they satisfy all the required inequalities, completing the proof of \Cref{thm:sum-admissible}. 

\subsection{A finite prefix and a telescoping tail}
As explained in \Cref{sec:weight}, the support identity suggests $h_d\sim C/d^2$ as $d\to\infty$.
For ease of computation, choose an integer $N\ge3$ and set
\begin{equation}
\label{eq:tail-h_d}
 h_d=\frac{C}{d(d-1)}\qquad\text{for }d>N \enspace .
\end{equation}
We leave $h_1,\ldots,h_N$ and $C$ unknown.
This choice gives an exact formula for the total sum:
\begin{equation}\label{eq:weight-objective}
 S:=\sum_{d\ge1}h_d=\sum_{d=1}^Nh_d+\frac CN,
 \qquad\text{and}\qquad W(d)=S-\frac Cd\quad\text{for }d>N \enspace .
\end{equation}
We do not assume that every admissible sequence, or an optimal one, must have the tail \eqref{eq:tail-h_d}.
We next construct an admissible sequence of this form.
As discussed before, we choose $C>2$ so that~$H_d$ grows faster than $\lambda(2h_d)$ for large $d$.
For the estimates below, we impose the slightly stronger
linear restriction
\begin{equation}
\label{eq:tail-threshold}
  C\ge2+\frac{4}{N(N+3)} \enspace .
\end{equation}
We still choose $C$ in the optimization, subject to this bound.
The bound ensures that the comparisons below hold for all large degrees, without needing to know the values of $H_d$ in advance.
\begin{lemma}
\label{lem:coltail}
  Suppose that \eqref{eq:tail-h_d} and \eqref{eq:tail-threshold} hold.
  For every $t\ge0$,
  \begin{equation*}
    \Phi(t)=\max_{1\le c\le N+1}
          \{\lambda(t+h_c)-A_c\} \enspace .
  \end{equation*}
  Moreover, $\sigma_d=H_d-\lambda(2h_d)$ is strictly increasing for $d\ge N+1$.
\end{lemma}
\begin{proof}
  For $d\ge N+1$, we have $h_d=C/[d(d-1)]$ and $h_{d+1}=C/[d(d+1)]$.
  Only the increment of~$H_d$ is needed:
  \begin{equation*}
    H_{d+1}-H_d=\frac{C(d+2)}{d(d+1)} \enspace .
  \end{equation*}
  Since $(d-1)(d+2)\ge N(N+3)$, \eqref{eq:tail-threshold} implies
  \begin{equation}\label{eq:tail-increment}
    \ln\frac{h_d}{h_{d+1}} =\ln\left(1+\frac2{d-1}\right)
     <\frac2{d-1} \le\frac{C(d+2)}{d(d+1)}=H_{d+1}-H_d \enspace .
  \end{equation}
  For $t\ge0$, \eqref{eq:lambda-difference} gives
  \begin{equation*}
    \lambda(t+h_{d+1})-\lambda(t+h_d)
    \le\ln\frac{t+h_d}{t+h_{d+1}}
    \le\ln\frac{h_d}{h_{d+1}}<H_{d+1}-H_d \enspace .
  \end{equation*}
  As $A_{d+1}-A_d=H_{d+1}-H_d$, the expression $\lambda(t+h_d)-A_d$ decreases for $d\ge N+1$. This proves the maximum formula.
  Applying the same comparison to $\lambda(2h_{d+1})-\lambda(2h_d)$ proves that $\sigma_d$ increases there as well.
\end{proof}

\subsection{The finite convex problem}
Choose integers $N\ge3$ and $M\ge N+2$.
The unknowns are $h_1,\ldots,h_N,C$; every later $h_d$ is given by~\eqref{eq:tail-h_d}.
In particular, all $H_d$ and $A_c$ appearing below are affine functions of these unknowns.
Consider the following finite problem:
\begin{equation}
\label{eq:finite-program}
\tag{P}
\begin{aligned}
 \text{minimize}\quad &S=\sum_{d=1}^Nh_d+\frac CN, &&\\
 \text{subject to}\quad
 &h_1\ge\cdots\ge h_N\ge\frac{C}{N(N+1)}, &&\\
 &C\ge2+\frac4{N(N+3)}, &&\\
 &\text{conditions \eqref{cond:first}, \eqref{cond:transfer}, and \eqref{cond:stop}}, &&\\
 &\lambda(2h_d)\le H_d+h_2
 &&2\le d\le N+1,\\
 &\lambda(2h_a)+\lambda(h_{a-1}+2h_a+h_c)\le H_a+A_c
 &&3\le a<M,\ 1\le c\le N+1,\\
 &\lambda(2h_M)+\lambda(h_c)\le H_M+A_c
 &&1\le c\le N+1 \enspace .
\end{aligned}
\end{equation}
The first two rows make the entire sequence positive and decreasing. The final row is a uniform bound for degrees beyond $M$.
Its value is imposed as a constraint in the optimization; we do not know $H_M$ in advance.

The objective is linear.
Conditions not involving $\lambda$ are linear.
Every other inequality is a sum of convex functions of affine arguments bounded above by an affine function.
Thus \eqref{eq:finite-program} is a convex optimization problem.
Its constraints contain no prescribed exponential base: a feasible objective value $S$ yields the base $e^S$ for our exponential upper bound.

\begin{proposition}
\label{prop:finite-feasible}
Every feasible solution of \eqref{eq:finite-program}, extended by \eqref{eq:tail-h_d}, gives an admissible sequence with finite sum $S=\sum_{d\ge1}h_d$.
Consequently, $J(Q)\le e^{F_h(Q)}\le e^{nS}$.
\end{proposition}
\begin{proof}
  Fix finite integers $N\ge3$ and $M\ge N+2$, and assume that \eqref{eq:finite-program} has a feasible solution.
  Extend its weights using \eqref{eq:tail-h_d}.
  The resulting weights are positive and decreasing, and satisfy \eqref{cond:first}, \eqref{cond:transfer}, and \eqref{cond:stop}, by the finite constraints.
  Their sum $S$ is finite by \eqref{eq:weight-objective}.
  For \eqref{cond:degree}, the finite comparisons give $\sigma_d\ge-h_2$ up to $N+1$, and \Cref{lem:coltail} extends this inequality to every larger $d$.

  For \eqref{cond:exceptional}, put $t_a=h_{a-1}+2h_a$.
  For $3\le a<M$, the finite constraints give $\lambda(t_a+h_c)-A_c\le\sigma_a$ for $c\le N+1$; \Cref{lem:coltail} extends this to all $c\ge1$.
  The last row of \eqref{eq:finite-program} says that
  \begin{equation*}
    \max_{1\le c\le N+1}\{\lambda(h_c)-A_c\}\le\sigma_M \enspace .
  \end{equation*}
  For $a\ge M$ and any $c\ge1$, monotonicity of $\lambda$ and \Cref{lem:coltail} therefore give
  \begin{equation*}
    \lambda(t_a+h_c)-A_c \le\lambda(h_c)-A_c
                         \le\max_{1\le j\le N+1}\{\lambda(h_j)-A_j\}
                         \le\sigma_M\le\sigma_a \enspace .
  \end{equation*}
  This is \eqref{cond:exceptional}.
  By \Cref{thm:admissible-potential}, $F_h$ is a branching potential.
  Hence \eqref{eq:branching-potential} gives $J(Q)\le e^{F_h(Q)}$. Since $F_h(Q)\le nS$, we obtain $J(Q)\le e^{nS}$.
\end{proof}

\subsection{The computed sequence}
By \Cref{prop:finite-feasible}, we may choose any integers $N\ge3$ and $M\ge N+2$, provided that we find a feasible solution of \eqref{eq:finite-program}.
The parameter $N$ determines how many weights we choose independently before using \eqref{eq:tail-h_d}.
The parameter $M$ determines how many degrees we check individually before applying the final uniform bound.
Increasing these parameters allows more flexibility, but also requires more computation.
We use $N=32$ and $M=512$.
These values give a feasible solution with $S<\ln2.81$, as certified below, while keeping
the computation small: there are only $33$ unknowns,
namely $h_1,\ldots,h_{32}$ and $C$.
In particular, the last constraint of \eqref{eq:finite-program} holds with $M=512$.
We only need one feasible solution giving the desired bound; the parameters need not be optimal.
We minimize $S$ numerically, allowing small positive margins in the constraints.
\footnote{\label{fn:weight-scripts}The weights were found and verified using the Python scripts
\texttt{find\_weights.py} and \texttt{verify\_weights.py},
respectively, available at \url{https://github.com/ebrahim-ghorbani/stable-matching}.
The first finds candidate weights using SLSQP with explicit derivatives; the second verifies the stated weights using exact rational arithmetic and certified bounds for exponentials.}
We then round the values to a fixed number of decimal places and check all constraints exactly, using fractions and rigorous upper and lower bounds for exponentials.
The resulting feasible solution has
\begin{equation}
\label{eq:chosen-C}
  C=2.013448663125,
\end{equation}
and the first $32$ weights listed in \Cref{table:h_d}.
These values and \eqref{eq:tail-h_d} specify the complete sequence.
Its total weight is
\begin{equation}
\label{eq:computed-S}
  S=1.03307089361665625,
  \quad\text{and}\quad e^S\approx2.809680831 \enspace .
\end{equation}
The values of $C$ and $S$ are exact.
\begin{table}
\centering\small
\begin{tabular}{rr|rr|rr|rr}\toprule
$d$&$10^{12}h_d$&$d$&$10^{12}h_d$&$d$&$10^{12}h_d$&$d$&$10^{12}h_d$\\\midrule
1&346573640280&9&19459568288&17&6773799446&25&3246293563\\
2&156726247027&10&16264151541&18&6096070881&26&3011434741\\
3&99061162159&11&14149746663&19&5509913479&27&2799883120\\
4&68322728671&12&12347392810&20&5000447702&28&2608984251\\
5&49770893583&13&10836070115&21&4555520046&29&2436248177\\
6&37763080994&14&9561451451&22&4165164703&30&2279535656\\
7&29565950277&15&8481331264&23&3821175296&31&2136999249\\
8&23740554322&16&7561379311&24&3516764526&32&2007039302\\\bottomrule
\end{tabular}
\caption{The first $32$ weights found by optimization and rounding.
Each listed integer equals $10^{12}h_d$.}
\label{table:h_d}
\end{table}

\begin{proof}[\bfseries Proof of \Cref{thm:sum-admissible}]
By direct computation (using \texttt{verify\_weights.py},
linked in \Cref{fn:weight-scripts}), we verify that the sequence specified by \Cref{table:h_d}, \eqref{eq:tail-h_d}, and \eqref{eq:chosen-C} is admissible and satisfies
  \begin{equation*}
    \sum_{d\ge1}h_d<\ln2.8097<\ln2.81 \enspace .
  \end{equation*}
The program checks the following inequalities:
  \begin{itemize}[itemsep=2pt]
    \item positivity and monotonicity of the prefix and its junction with $h_{33}$, together with \eqref{eq:tail-threshold};
   \item conditions \eqref{cond:first}, \eqref{cond:transfer}, and \eqref{cond:stop};
   \item $\lambda(2h_d)\le H_d+h_2$ for $2\le d\le33$;
   \item $\lambda(2h_a)+\lambda(h_{a-1}+2h_a+h_c)\le H_a+A_c$ for $3\le a<512$ and $1\le c\le33$;
   \item $\lambda(2h_{512})+\lambda(h_c)\le H_{512}+A_c$ for $c = 1,\hdots,33$.
  \end{itemize}
  All comparisons use exact rational arithmetic and rigorous upper and lower bounds for exponentials.
  For instance, an inequality $\lambda(x)+\lambda(y)\le z$ is checked in the equivalent form
  \begin{equation*}
    (1-e^{-x})(1-e^{-y})\ge e^{-z} \enspace .
  \end{equation*}
  \Cref{prop:finite-feasible} therefore proves admissibility for all degrees, including those beyond the finite checked ranges.

  We obtain the exact value of $S$ in \eqref{eq:computed-S} by adding the first $32$ weights in \Cref{table:h_d} and $C/32$, which is the sum of all remaining weights.
  The same certified bounds for exponentials give $28097\,e^{-S}>10000$, which is equivalent to $S<\ln2.8097$.
\end{proof}

Exact feasibility is the numerical input needed by the proof.
We make no claim that the candidate is an exact optimizer of \eqref{eq:finite-program}; proving optimality is unnecessary for the upper bound.

\noindent
\paragraph{Acknowledgments.} The second author sincerely thanks Ildik{\'o} Schlotter for introducing the problem, and for discussions substantiating the entropy-based approach of Palmer and P{\'a}lv{\"o}lgyi.

\bibliographystyle{alpha}
\bibliography{bibliography}
\end{document}